# Understanding data analysis aspects of TMS-EEG in clinical study: a mini review and a case study with open dataset

Hua Cheng[①]

---

[①] trernghwhuare@aliyun.com; trernghwhuare@hotmail.com. (extra: leoloverory@outlook.com)

**Understanding Data Analysis Aspects of TMS-EEG In Clinical Study: A Mini Review**

# Abstract

Concurrency of transcranial magnetic stimulation with electroencephalography (TMS-EEG) technique is a powerful and challenging methodology for basic research and clinical applications. Aspects considered in experiments for effective TMS-EEG recordings and analysis, including artifact management, data analysis and interpretation and protocols. This review offers an extensive insight of TMS-EEG methodology in experimental and computational procedures.

# Introduction

Transcranial magnetic stimulation (TMS) produces non-invasive brain stimulation to probe neurophysiological processes within the brain[1]. TMS pulse initiated by flowing an intense current through the TMS coil windings. The current inducing an E-field is a time-varying magnetic field that penetrates the scalp and skull unimpeded. And the eddy currents induced in the brain can depolarize neurons. E-field along neurites changes rapidly at cortical neurons that have axonal bends or other geometrical inhomogeneities or endings. With short pulse duration of 1-3 T in strength and a rise time of about 50-100 μs, TMS has temporal resolution of sub-milliseconds which allows for real-time modulation of the brain. Superficial cortical layers simulated more strongly than deeper layers as the results of magnetic fields attenuate rapidly with distance and the induced E-field approaches zero at the center of the head. But when applying adequate stimulation intensity (SI) by TMS, action potentials evoked locally may propagate along anatomical connections across cortical layers within the same cortical column and to other cortical and subcortical regions, and may result in the activation of an entire network[2].

Electroencephalogram (EEG) studies the electrophysiological dynamics in brain non-invasively with millisecond temporal resolution and centimeters spatial resolution *via* measuring differences in electrical potential of postsynaptic potentials synchrony rather than action potentials between electrodes placed on the scalp[2].

Compared to other neuroimaging techniques like fMRI, near-infrared spectroscopy (NIRS) and PET that can record TMS evoked neuro activity, EEG is the most successful and commonly used combination has been with EEG due to its inexpensive and simplicity to combine online with TMS[2]. TMS-EEG data derived from EEG responses to TMS can be used as a neurophysiological marker of excitability or connectivity in cortex. TMS-EEG is capable of manipulating and investigating brain rhythms by measuring the impact of a TMS pulse on EEG and associated behavioral effects further investigated in the frequency domain.

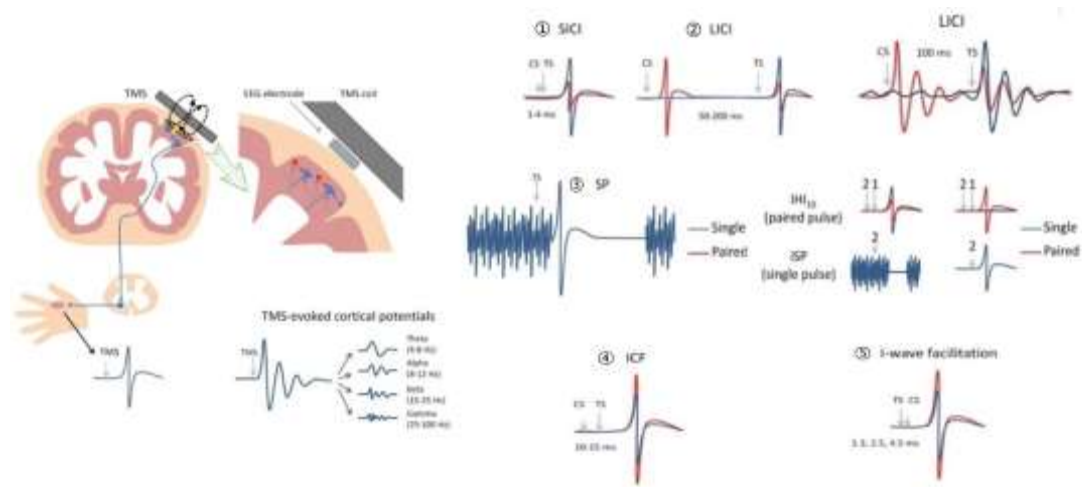

Fig. 1 TMS over cortex and TMS paradigms assessing various inhibitory and excitatory neuronal populations. ①Short-interval intracortical inhibition (SICI) involves comparing MEP amplitude of a single, suprathreshold test stimulus (TS) to a paired-pulse condition with a subthreshold conditioning stimulus (CS) and suprathreshold TS at 1-4-ms intervals. ②Long-interval cortical inhibition (LICI) involves comparing a suprathreshold with a paired-pulse suprathreshold CS and TS at 50-200-ms

intervals. ③Silent period (SP) involves measuring the duration of absent muscle activity following a single, suprathreshold TS given during a muscle contraction. ④Intracortical facilitation (ICF) involves comparing a suprathreshold TS with a paired-pulse subthreshold CS and suprathreshold TS at 10-15-ms intervals. ⑤I-wave facilitation involves comparing a suprathreshold TS with a paired-pulse condition in which a subthreshold CS follows the TS at specific intervals of 1.3, 2.5, and 4.5ms. Adapted and modified from *N. C. Rogasch, et al.*, 2014[3].

### *TMS-evoked EEG potentials (TEPs)*

TMS-evoked EEG potentials (TEPs) are the EEG responses to TMS averaged in the time domain[4]. Cortical excitatory (glutamatergic) and inhibitory (GABAergic) neurotransmitter systems activations create separate components or peaks that construct TEP. TMS of both the motor and the frontal cortices of healthy adults generally elicits a sequence of TEP components (see *Fig.*2-A)or positive (P) and negative (N) peaks at around milliseconds of 30 (P30), 45 (N45),60 (P60), 100 (N100),180(P180), and 280(N280) [5].It is believed that peaks within the first 30 ms (P30) reflect excitatory neurotransmission[6-8]. N45 and N100 peaks (see *Fig.*2-B, C) are associated with $GABA_A$ and $GABA_B$ receptor-mediated neurotransmission, respectively[9]. Later peaks have been linked to the balance between glutamatergic excitatory and GABAergic inhibitory neurotransmission[6,7,10,11].

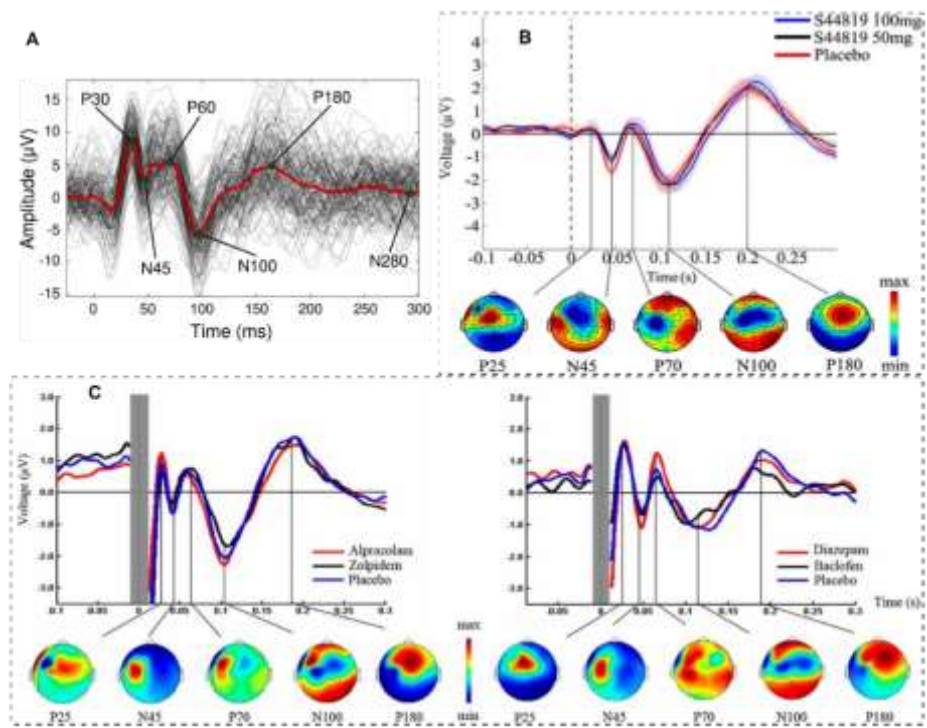

Fig. 2 TMS-evoked EEG potentials. (A)Typical TEPs(P30-N45-P60-N100-P180-N280) in a healthy young adult. Adapted from *E. Kallioniemi, et al.*, 2022[5].(B) Grand average TEPs after single-pulse TMS of M1 at baseline in the three different drug conditions of selective α5-$GABA_A$R antagonist. Adapted from *G. Darmani, et al.*, 2016[9]. (C) grand average TEPs after single-pulse TMS of M1 at baseline in conditions of subunit-containing $GABA_A$Rs (left) and specific $GABA_B$R agonist(right). Adapted from *I. Premoli, et al.*, 2014[7].

Many aspects of TEPs remain to be better characterized such as their morphology and physiology in different non-motor regions, test-retest reliability of peaks measured in several non-motor regions, the specific origin of each of the TEPs peaks, the exact involvement of somatosensory and auditory evoked responses in specific cortical regions and the involvement of subcortical structures in their generation.

**Electrodes of interest (EOI)** quantify TEPs amplitude and latency over subsets of electrodes, is one of the most common approaches. Electrodes selection in EOI analyses can be approached in priori or using data-driven methods. Data are presented as a waveform of varying amplitude as a function of time. Scalp voltage distributions visualize and quantify the spread of activity at selected time points across the cortex in ROI. EOI method is particularly relevant when there is a clear a priori hypothesis on the location of the expected brain response evoked by TMS. This method is not optimal when there is a big TMS artifact as it can mask the TEP peaks.

**Local mean field power (LMFP) / cortical evoked activity (CEA)** measures the area under the curve (*i.e.*, the integral) of the rectified signal or standard deviation (root mean square) across specific EOI at a given point in time corresponding to TEP peaks. LMFP/CEA is an alternative approach in measuring TEP peak amplitudes and latencies. LMFP/CEA does not present an obvious main peak and ignores the polarity of the signal, but it takes into account the width as well as the peak of the evoked activity. The LMFP method is relevant when there is an a priori hypothesis related to an expected change in brain activity that is localized and not related to a specific TEP peak (see *Fig.*3).

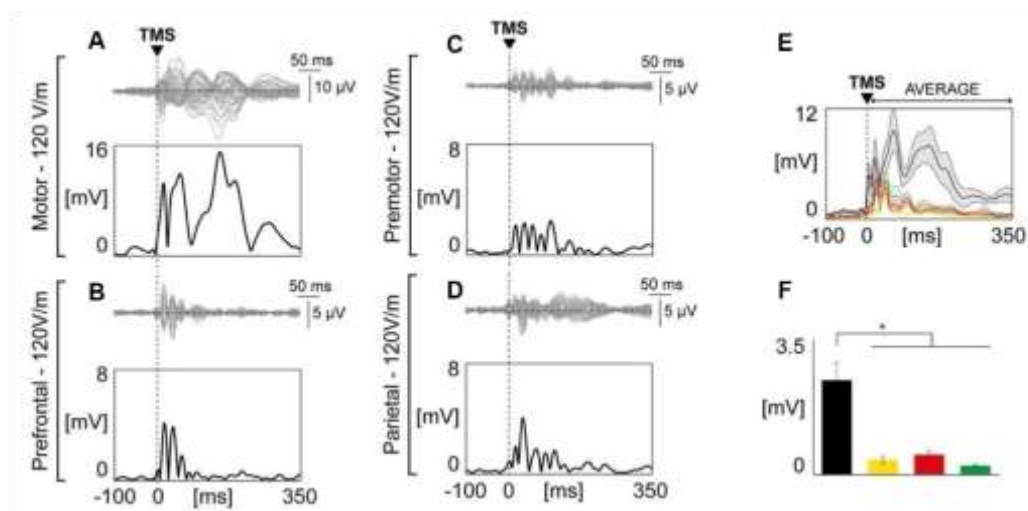

Fig. 3 different cortical areas employing LMFP. Butterfly plots of all channels with corresponding LMFP in area of (A) motor; (B) prefrontal; (C)premotor; (D)parietal. (E) Grand-average of LMFP for each stimulated area. Thick traces indicate the grand-average LMFP across subjects. Responses recorded after the stimulation of different cortical areas are color coded as follows: motor in black, prefrontal in yellow, premotor in red, parietal in green. (F) the LMFP values averaged between 8 and 350 ms post-TMS for each stimulated area. Adapted and modified from *M. Fecchio, et al.*, 2017[12].

**The global mean field power/amplitude (GMFP/GMFA)** measures the impact of the TMS pulse on activity evoked across all electrodes, which is the averaged signal of TMS activity over the entire surface of the head, or the standard deviation (root mean square) across electrodes at a given point in time.

$$GMFP(t) = \sqrt{\frac{\left[\sum_i^k \left(V_i(t) - V_{mean}(t)\right)^{\wedge}2\right]}{k}} \quad (1)$$

Where $t$ is time, $t$ is the number of channels, $V_i$ is the voltage in channel $i$ averaged across subjects and $V_{mean}$ is the mean of the voltage in all channels.

GMFP analysis is the method of choice when there is no a priori hypothesis with regards to local activity, but rather when the goal is to explore global brain activity following the TMS pulse[13] (*Fig.*4).

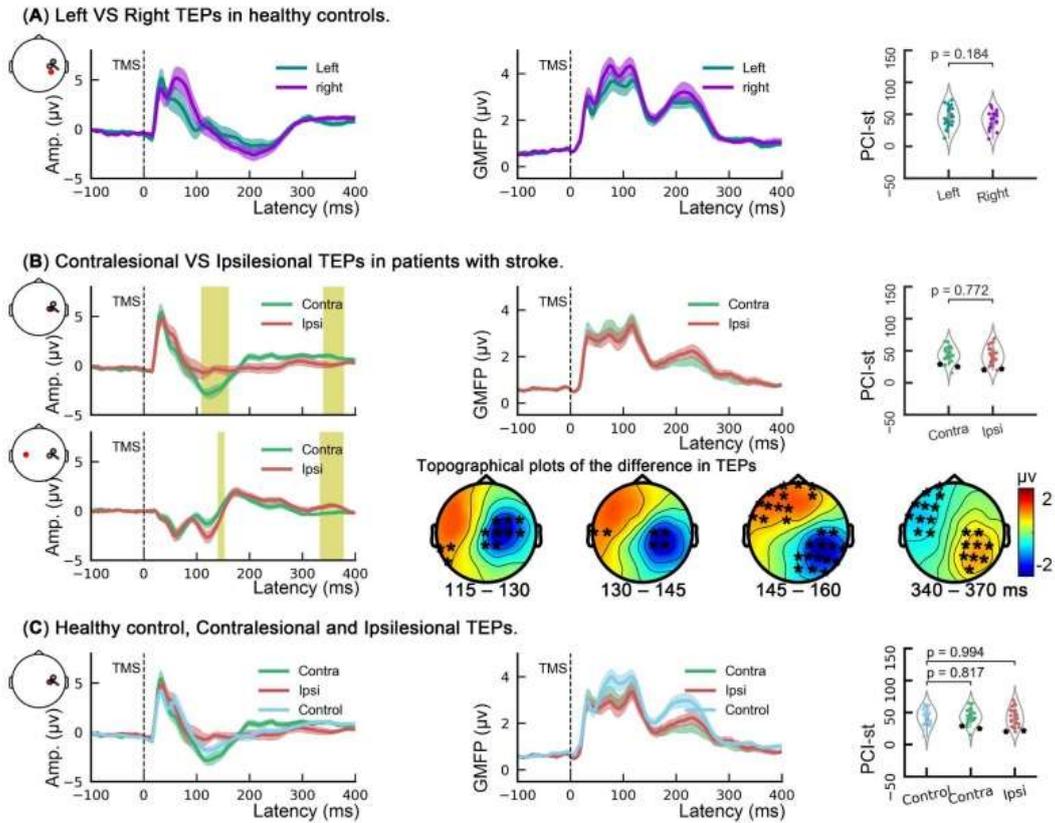

Fig. 4 TEPs Analyses. (A) TEPs comparison of the left and right in healthy controls: the TEP amplitudes of a representative channel, the GMFP of the TEPs and the comparison of left and right PCI-st. (B) TEPs comparison of the contralesional and ipsilesional in patients with stroke: the TEP amplitudes of representative channel located at the right and left hemispheres. Yellow rectangles indicate time windows in which significant differences between the ipsilesional and contralesional hemispheres were found. These differences are shown in four topographies with the black asterisk (*) represents significant clusters. And plots of The GMFP and PCI-st plots. (C) TEPs comparison in the patients with stroke and healthy controls. Adapted and modified from *Z. Bai, et al.*, 2023[14].

*TMS-induced EEG oscillations*

**TMS induces oscillations (TIOs)** specific to the brain area that can be quantified with EEG frequency domain analyses[5]. Resting-state EEG power was classified TMS-related oscillation for discrete frequency bands, *i.e.*, $\delta$ (2-4 Hz), $\theta$ (4-7 Hz), $\alpha$ (8-12 Hz), $\beta$ (13-30 Hz) and $\gamma$ (30-45 Hz) frequency bands. Time-frequency decomposition of the TMS-EEG signal reveals TIOs' typical profile following M1 stimulation is characterized by an early increase of $\delta$, $\theta$, $\alpha$ and $\beta$ band power up to 200 ms, followed by $\alpha$ and $\beta$ suppression (often termed de-synchronization) with a final increase in $\beta$ power. The occipital cortex TMS evokes $\alpha$ oscillations, parietal cortex TMS evokes $\beta$ oscillations, and frontal cortex TMS evokes fast $\beta/\gamma$ oscillations (see *Fig. 5*)[15].

TMS mostly synchronizes pre-existing and ongoing oscillations instead of eliciting new neural responses[16]. The induced time and phase-locked oscillations create a TEP, while the non-phase-locked responses induced that average out in the TEP can be seen with specific signal-processing methods[17].

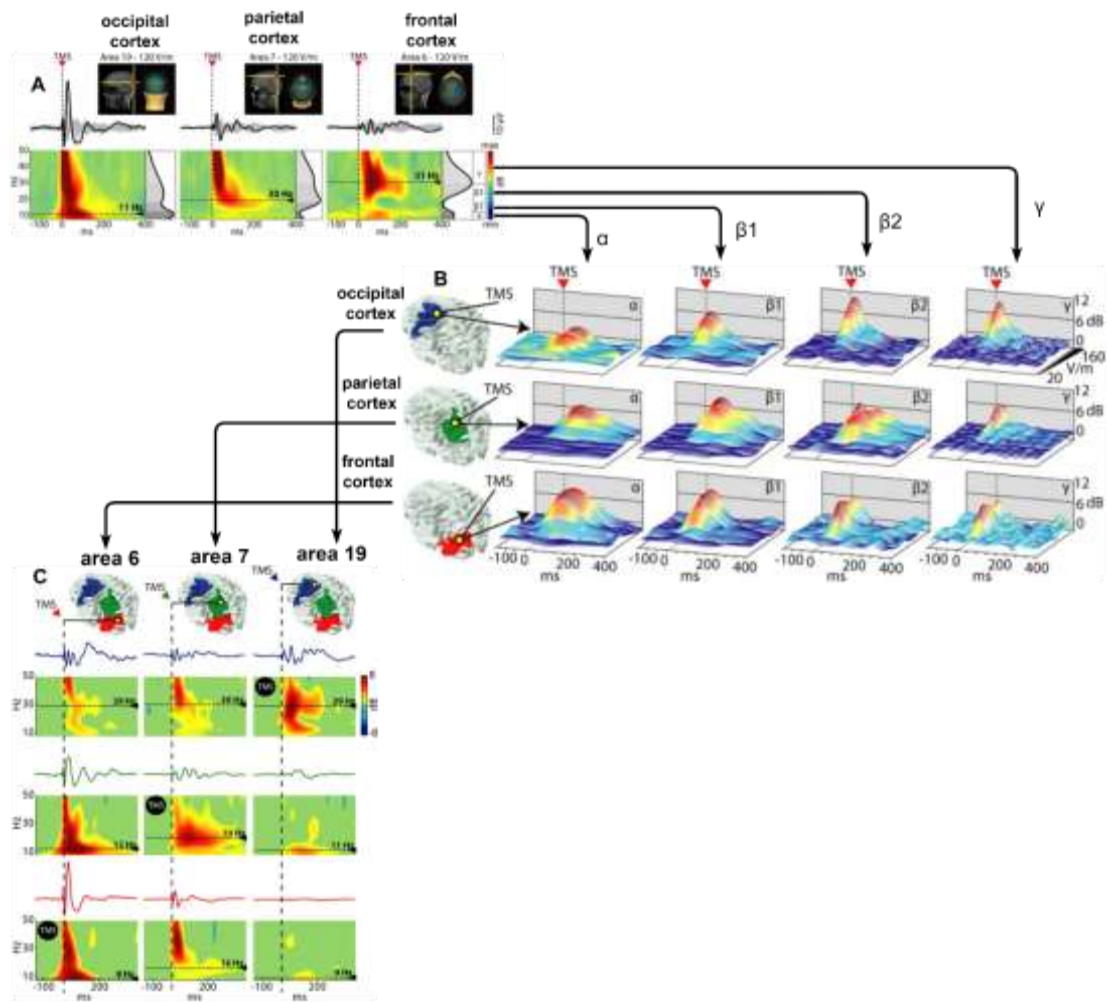

Fig. 5 TMS induces oscillations. (A) TMS elicited early $\gamma$ components immediately followed by prominent $\alpha$-band oscillations after occipital stimulation, $\beta$-band oscillations after parietal stimulation, and fast $\beta/\gamma$ oscillations after perturbation of frontal cortex. (B) Different natural frequencies in different cortical areas are not attributable to different stimulation intensities. EEG frequency bands ($\alpha$: 8-12; $\beta$1:13-20; $\beta$2: 21-29; $\gamma$: 30-50) (C) The natural frequency is a local property of individual corticothalamic modules. Adapted and modified with the permissions of *M. Rosanova, et al.*, 2009[15].

## Time-frequency approach

**Time-frequency representation (TFR) approaches**, such as, wavelet transforms (WT), short-time Fourier transform (STFT), extract the frequency and amplitude of cortical oscillations over time. The one focused on evoked oscillatory response (EOR), while the one accounts for the so-called induced oscillatory response (IOR) but actually better characterized by the definition of total oscillatory response (TOR) is the two approaches are generally used[18]:

**EOR** involves applying the time-frequency decomposition to the data averaged across trials (*e.g.*, the TEP) and returns information only on phase-locked oscillations following TMS (*i.e.*, evoked oscillations). **TOR** involves applying the time-frequency decomposition to individual trials, and therefore captures both the phase-locked and non-phase locked oscillations following TMS (*i.e.*, evoked and induced oscillations)[13].

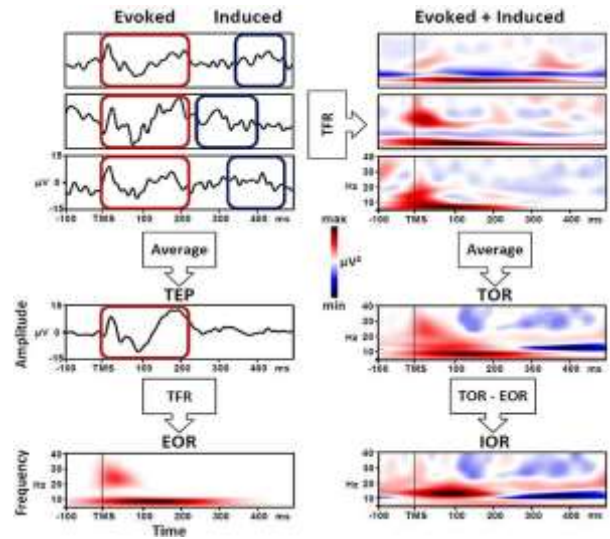

Fig. 6 Evoked oscillatory response (EOR), induced oscillatory response (IOR) and total oscillatory response (TOR) triggered by TMS pulse. **Left panel**: EOR is the time-frequency representation (TFR) of the average across all single cortical responses to TMS pulse (TEP). **Right panel**: TOR is the average of the TFR of each single response to TMS pulse, and includes EOR and IOR. To isolate pure IOR, the EOR must be removed from the TOR. Adapted from *M. C. Pellicciari, et al,*2017[18].

## TMS-EEG methodology and artifact correction

A basic TMS-EEG system consists of a TMS stimulator and coil, a TMS-compatible EEG amplifier and TMS-compatible EEG electrodes[5].

**Signal-to-noise ratio (SNR)** The SNR depends on the square root of the number of trials, provided that the meaningful signal and noise remain similar from trial to trial. The total SNR of averaged responses, such as TEPs, is,

$$SNR = \left(\frac{S}{N}\right) \times \sqrt{T} \qquad (2)$$

Where $S$ is the size of the signal, $N$ is the size of the noise on a single trial, and $T$ the number of trials. The SNR on a single trial is defined as $S/N$ (the signal divided by the noise)[2].

**TMS threshold determination** Thresholds can be determined by measuring motor threshold (MT), phosphene threshold (PT), TEP amplitude, or induced E-field[2].

*Artifacts in TMS-EEG signals*

**TMS-EEG artifacts**

The major hindering the identification of TMS-EEG responses are physiological and non-physiological artefacts that interfere with the measured EEG signal. TMS coil generating electromagnetic field produces artifact in concurrent application of TMS and EEG are several orders of magnitude larger than electrophysiological activity of the brain recorded by the EEG, which resulted in saturation of EEG amplifiers. Common artifacts in TMS-EEG recordings include

1) common EEG artifacts. EEG artifacts arise from environmental noise (*e.g.*, power line) and physiological noise (*e.g.*, Eye blinks, cranial muscle twitch, auditory responses to the coil click, and SEPs, are all physiological but unwanted signals that can be induced by the TMS pulse).
2) TMS-related artifacts. The coil inevitably contacting with the electrodes, movements of EEG sensors, the pressure of the coil on the electrodes, the magnetic field applied on the electrode, the electrode-skin interface, as well as the capacitor recharge in TMS stimulators will also contribute to the production of TMS-related artifacts in the signal. TMS-induced decay artifact Offline procedures for artifact removal is a large positive shift in the signal that linearly recovers within up to 50 ms. Using TMS-compatible recordings and off-line artifacts correction, the decay artifact can recover within 10-12 ms allowing to measure early latency TMS-evoked potential. For artifacts reduction, using EEG electrodes designed for TMS applications, appropriate skin preparation to lower signal impedance under the coil for reducing direct contact with electrodes and the electrode wires re-orientation perpendicular to the stimulating coil can also help minimize the TMS-decay artifacts.

Some confounding factors secondary to the TMS pulse that should be reduced by adopting specific strategies. They are,

1) the TMS pulse inducing loud clicking noise (100-120 dB) can cause an auditory-evoked potential.
   Wearing sound protective headphones and/or playing white noise in earphones is typically used to maximally reduce this artifact.
2) the TMS pulse activating sensory afferents results in a tapping sensation on the scalp that can induce a somatosensory-evoked potential.
   Using a thin layer of foam under the coil may help attenuating this effect.
3) the TMS pulse also produce facial muscle activation and time-locked blinks.
   Making sure that TMS elicits strong initial cortical responses at the stimulation site. Unlike SEPs, TEPs are specific for the stimulation parameters, like site, intensity, orientation and are characterized by even larger responses upon loss of consciousness. Most important, intracranial electrical stimulation can replicate specific changes across states in the time-frequency features and overall complexity of the responses to TMS without eliciting any sensory percept both at early and late latency. To the extent and morphology of the impacted cortex, eliciting prominent cortical responses to TMS depends on stimulation intensity, coil orientation and design. Thus, develop and apply real-time standardized data visualization tools during the experimental procedures is important to ascertain the amplitude of early TEP components. It's useful that using a sham condition to control sensory-related confounding factors in experiments where TMS is aimed at exploring subtle changes occurring in specific, local circuits[13].

**Offline procedures for artifact removal**

Blind source separation (BSS) method unmixes original source signals from their intermixed observations without prior knowledge of the mixing algorithm or source signals.

$$S = WX; S \in [m,t]; W \in [m,n]; X \in [n,t] \quad (3)$$

Where $n$ is the number of channels and $m$ the number of independent components; $t$ represents the time course.

BSS removing TMS related artifacts help contributing to TMS-EEG development, includes,

1) Independent component analysis (ICA), which is assumed that EEG signals originate from temporally and spatially independent sources and can be modeled as a linear combination of cortical and non-cortical sources with independent time courses. But if the assumption of independence is not valid, then ICA may not separate the artifacts correctly. ICA-based artifacts correction optimization follows analysis pipeline (*Fig.*7).

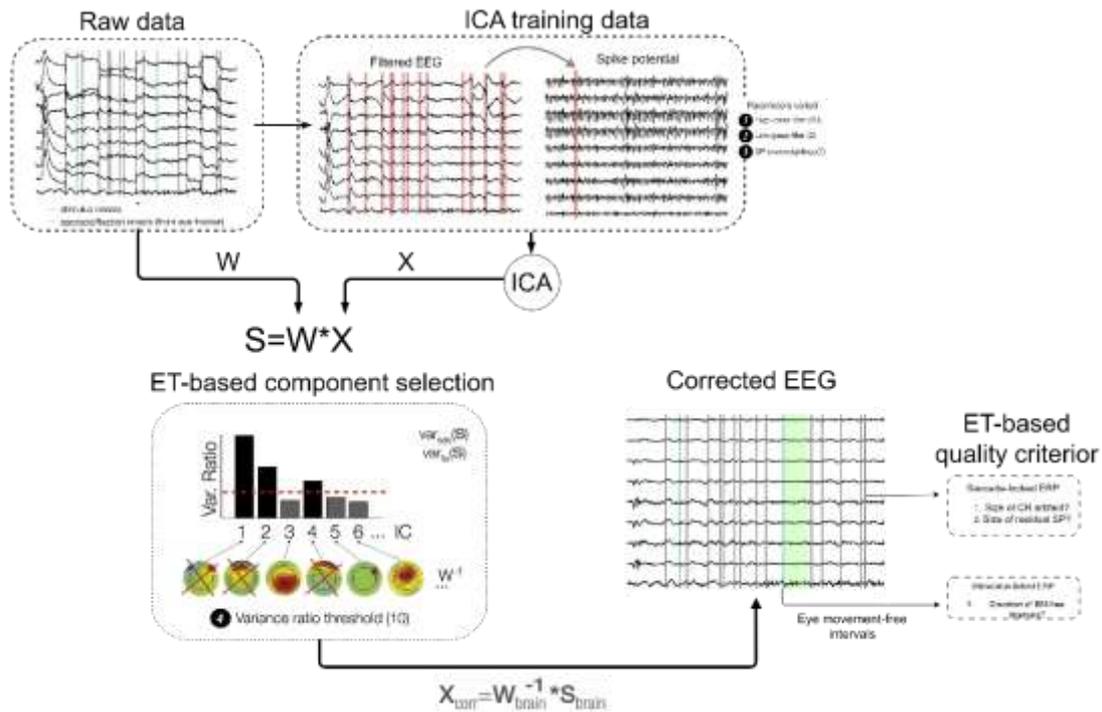

Fig. 7 Schematic workflow to determine parameters for optimized ICA training and component classification. Adapted from *O. Dimigen,* 2020[19].

2) Principal components analysis (PCA), applied to remove eye blink artifacts in EEG signals initially, is based on a linear combination of orthogonal principal components. PCA linearly transforms a set of input data channels into an equal number of linearly-uncorrelated variables and reduce the dimensionality by orthogonal rotation, a preprocessing step of ICA.

The artifact corrected with ICA and PCA using data from all the electrodes to smooth the signals, while per electrode artifact correction may be effective when the intensity of the TMS related artifacts is locally concentrated.

# TMS-EEG data analysis

When analyzing TMS-EEG data, parameters and protocols should be carefully chosen and controlled for when designing TMS-EEG experiments. Three types of parameters that can be selected in TMS-EEG experiments are parameters of input as TMS parameters, output as EEG parameters and brain state parameters (*Fig.* 8).

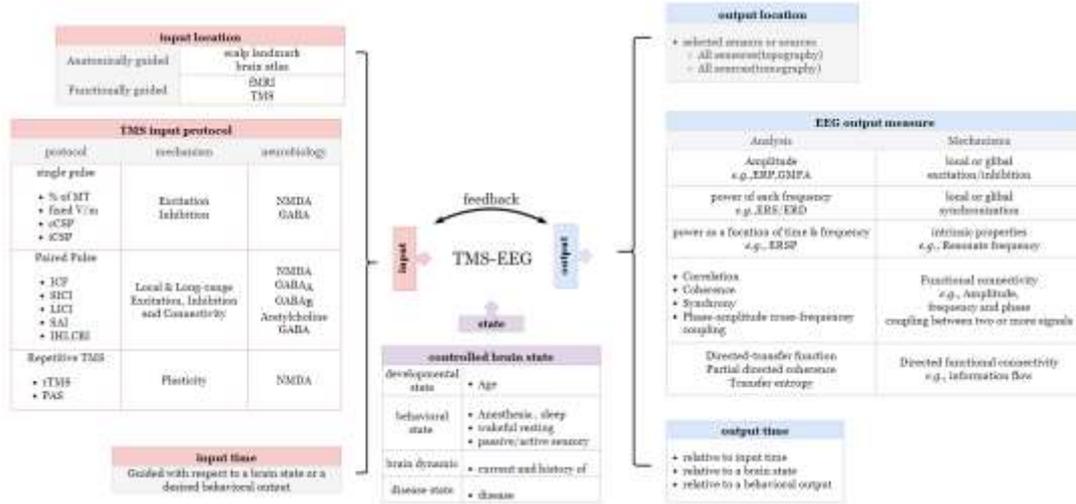

Fig. 8 A system diagram guiding the design of TMS-EEG studies. Adapted and modified from *F. Farzan, et al*, 2016[20].

While many steps for standard EEG preparation can be applied to TMS-EEG[21], additional steps are required to minimize the impact of confounding factors and artifacts introduced by TMS. specific requirements for TMS-EEG preparation, *e.g.*, very low impedances (<5 kΩ), positioning of reference and ground electrodes far from the stimulation target, proper selection of the EEG amplifier settings (hardware filtering bandwidth, sampling rate, amplitude resolution).

The TMS-EEG data construction resulted in a three-dimensional tensor representing a time varying spectrum of all channels: channel (or space) × frequency × time[22].

Assuming a 3D tensor $W$, each component comprises three matrices ($A$, $B$ and $C$), as,

$$w_{ijk} \approx \sum_{r=1}^{n} a_{ir} \cdot b_{jr} \cdot c_{kr} \qquad (4)$$

Where $w_{ijk}$ is an element in the tensor $W$, which is approximated by the summation of N rank-1 components which are the outer product of $a_r, b_r, c_r$, where $a_{ir}$ is an element in the matrix $A$ which contains the profiles of the extracted components along the first dimension (channel or space) in its columns $a_r$. Likewise, $B$ and $C$ contains the estimated components along the second (frequency) and third (time), respectively[22].

EEG analysis involves quantification of EEG signals in terms of amplitude, frequency, phase, the interaction between these attributes, the direction of information flow, and the dynamics of EEG topography, chronometry or tomography. Extracted from one or more sensors or sources, the EEG features can be described relative to the time of TMS application or change in the brain state.

## Linear TMS-EEG analysis

EEG analysis is often based on the assumption that the EEG signal represents a linear dynamical system. As the equation,

$$Y = B + A + N = LS + L_A S_A + L_N S_N \qquad (5)$$

Where $Y$ represents the EEG recorded signals, $B$ represents the brain signals of interest, $A$ represents the sum of the artifacts, and $N$ represents the noises that contaminate the recorded data. When $B = LS$, $B$ equals a product of two matrices $L$ and $S$. Where $L$ is the lead field or mixing matrix whose entry $L_{i,j}$ determines the sensitivity of channel $i$ to the source $j$, and $S$ is the source matrix whose entry $S_{j,t}$ denotes the amplitude of the source $j$ at a time $t$. Meaning, the elements $A_{i,t}$ and $N_{i,t}$ of matrices $A$ and $N$ add artifacts and noise to the recorded signal $Y_{i,t}$. $L_A, S_A, L_N,$ and $S_N$ represents the artifact-mixing-, artifact-signal-, noise-mixing-, and noise-signal matrices, respectively[2].

When considered a linear dynamical system, the EEG signal can be decomposed into Fourier series, *i.e.*, Sine waves described by amplitude, frequency, and phase. In this model, amplitude represents the maximum vertical peak of the Sine wave (unit of $\mu V$), frequency is the number of complete cycles per second (unit of Hz), and phase describes the time point position with respect to the beginning of the Sine wave (unit of radian or degrees, ranging from -180° to 180°). To obtain the frequency and phase component, the EEG time series is multiplied by a transfer function, such as Fast Fourier Transform (FFT) or discrete wavelet transforms. In this procedure, a complex number is identified that can be used to compute the instantaneous power (proportional to the square of the maximum amplitude that the signal could reach) and phase of the signal.

## Non-linear TMS-EEG analysis

The EEG signal can also be considered as a non-linear, stochastic or deterministic, and dissipative dynamical system. In a non-linear dynamical system, described by non-linear equations, a small change in initial conditions may cause a large effect. In non-linear EEG analysis, chaos theory may be applied to reconstruct an attractor from the EEG time-series. The attractor is described by its dimension, Lyapunov exponents, and entropy. The non-linear EEG analyses were employed to describe non-linear synchronization between brain regions and network nodes. EEG analyses can be grouped into general categories of reactivity and connectivity analysis.

1) The aim of reactivity analysis is to characterize the regional or global brain response to an event or change in brain state. In these analyses, EEG signals are often characterized by

**Temporal analysis** Identifying time domain features including latency and amplitude of event-related potentials (ERP)s or evoked potentials (EP)s and Global Mean Field Amplitude (GMFA).

**Frequency analysis** Decomposing the time domain signals into frequency sub-bands including $\delta$ (~1-3 Hz), $\theta$ (~4-7 Hz), $\alpha$ (~8-12 Hz), $\beta$ (~13-28 Hz), and $\gamma$ (~>30 Hz) oscillations, and identifying outcome measures such as evoked and induced power, relative and absolute power, or event-related synchronization (ERS) or desynchronization (ERD).

**Time-frequency analysis** Performing spectral decomposition using a sliding time window to calculate the change in power of each frequency as a function of time, thereby, revealing time and frequency domain information and identifying outcome measures including event-related spectral perturbation (ERSP).

**Phase analysis** Identifying the phase of the EEG signal at a specific time point or relative to an event.

2) The aim of connectivity analysis is to describe how two or more functional units, such as two or more brain regions, network nodes/hubs, or brain dynamics (*e.g.*, oscillatory activity) interact, such as function in "synchrony," to form a larger-scale functional unit that underlies a specific brain-state. Connectivity techniques fall within two broad classes.

**Non-directed connectivity analysis** Measuring without quantification of the direction of information flow, including correlation, coherence or synchrony. These describe the relationship between signals recorded across the sensors (or sources), and/or across trials, by quantifying the interaction between signal attributes such as amplitude, frequency, and phase. Numerous connectivity and network dynamic metrics can be realized by quantifying the interaction between EEG features across brain regions.

**Directed connectivity analysis** Measures includes directed transfer function and partial directed coherence based on the Granger causality principle. For example, Directed Transfer Function (DTF) allows to determine the sources localization and the EEG activity propagation direction. defined by the Akaike information criterion (AIC) as follows,

$$AOC(d) = 2 \times \log(\det(V)) + \frac{2kd}{N} \tag{6}$$

Where $V$ is the noise variance matrix, $N$ is the window size, $d$ is the model order, and $k$ is the number of EEG channels. Model quality might fit if the conditions are satisfied: $k \times d < 0.1 \times N$ [23].

These measures can capture the direction of information flow, rather be complex computationally, and had been applied to EEG data recently. Notably, the validity and reliability of EEG markers of functional connectivity should be examined against simulated data. Studies suggest that some connectivity analyses are confounded by the effects of volume conduction and are sensitive to the methods of temporal filtering and source reconstruction[20].

### *Toolboxes for TMS−EEG data analysis*

Common EEG analysis software toolboxes used *Fieldtrip* and *EEGLAB*, combine their use with custom-written scripts on the *Matlab* platform. This complexity and the lack of a common "gold standard" analysis approach currently limit the implementation of TMS-EEG laboratories in clinical settings that do not have a strong expertise in scripting/coding, Moreover, this also contributes to the current heterogeneity in techniques employed and restricts generalization of results between studies. As such, the very recent publication of two open-source analysis approaches for TMS-EEG pre-processing, *i.e.*, *TESA* software and *TMSEEG* toolbox, as well as functionality within the *FieldTrip* toolbox, is an important step towards a standardization of TMS-EEG analysis procedures and will definitively facilitate the development of the field in the upcoming years[13].

# TMS-EEG protocols in clinical study

TMS can be used to measure various parameters in motor cortex and, allowing us to evaluate different aspects of cortical excitability. The threshold for producing an MEP in resting muscle reflects the excitability of a central core of neurons, which arises from the excitability of individual neurons and their local density. As it can be influenced by drugs that affect sodium and calcium channels, threshold must indicate membrane excitability.

## *TMS protocols for assessment of cortical inhibition and excitation*

Various single-pulse and paired-pulse TMS-EMG techniques empower the evaluation of both inhibition and excitation in M1. These methodologies have also been integrated into TMS-EEG experiments and extended to areas beyond M1, thereby broadening the scope of research for investigating excitatory and inhibitory processes across the cortex. The principle underlining in such TMS-EEG studies is the intracortical inhibitory or excitatory processes indexed by a change in EMG amplitude related MEPs could be quantified *via* TEPs and TMS-evoked measures such as evoked cortical oscillations.

The most frequently used protocols and metrics include evaluation of motor threshold, ipsilateral cortical silent period (iCSP), contralateral cortical silent period (cCSP). Paired-pulse measures intracortical facilitation (ICF), short interval intracortical inhibition (SICI), long interval intracortical inhibition (LICI), interhemispheric inhibition (IHI), cerebellocortical inhibition (CBI), short-latency afferent inhibition (SAI).

These measures investigate the integrity of a cascade of fast- and slow-acting excitatory and inhibitory processes, occurring either within local cortical circuitry or involving long-range cortico-subcortical feedback loops[20].

**Long-interval intracortical inhibition (LICI) and Cortical silent period (CSP)**

LICI and CSP are thought to be as indices of $GABA_B$ receptor inhibition. **LICI** is obtained when two suprathreshold stimuli are applied at intervals between 50 and 200 ms, and is thought to reflect $GABA_B$ receptor ($GABA_BR$) mediated neurotransmission. Application of LICI to the motor cortex results in attenuation of $\delta$, $\theta$, and $\alpha$ oscillations, whereas $\beta$ and $\gamma$ oscillations were significantly inhibited in DLPFC[24](*Fig*.9).

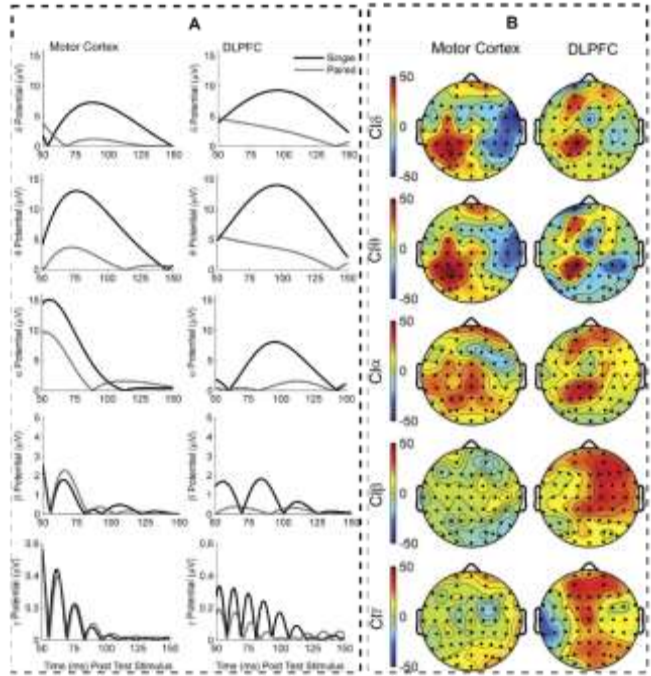

Fig. 9 Cortical inhibition (CI) measured by TMS-EEG through LICI and CSP paradigms. (A) TMS evoked cortical oscillations following the application of LICI to the motor cortex and DLPFC. (B) Topographic illustration of modulation of cortical oscillations following application of long-interval cortical inhibition (LICI) to the left motor cortex and dorsolateral prefrontal cortex (DLPFC). Adapted and modified from F. Farzan, et al.,2010[24].

The CSP is obtained when a TMS pulse is administered during target muscle's tonic contraction and serves as an indicator of cortical activity inhibition, most likely reflecting $GABA_BR$-mediated neurotransmission. $GABA_B$ergic-mediated inhibition in M1 was determined by the duration of CSP[25].

**Short interval intracortical inhibition (SICI) and intracortical facilitation (ICF)**

In SICI, a subthreshold conditioning stimulus (CS) inhibits a suprathreshold test stimulus (TS)-elicited MEP at inter-stimulus intervals (ISIs) of 2-3 ms while longer ISIs (7-30 ms) produce facilitation of MEPs, like ICF, for instance. SICI protocol is associated to $GABA_A$ receptor ($GABA_AR$) activity, while the ICF produce excitation associated to both $GABA_AR$ and NMDAR[13]. SICI and ICF led to inhibition and facilitation modulation of P30 and P60 TEP amplitude with TMS at M1, while P60 was bidirectionally modulated by SICI and ICF in the same manner whereas P30 was absent when DLPFC stimulation[26] (see *Fig.*10). An increase of amplitude of N100 by the SICI paradigm, N45 amplitude increased and N100 amplitude decreased by ICF indicated age-related alterations of excitatory and inhibitory functions in the prefrontal cortex in healthy adults[27] (see *Fig.* 11). Modulation of P60 by SICI and N100 by ICF may be associated with prefrontal $GABA_A$ and glutamatergic dysfunctions, in the expression of symptoms of schizophrenia[28](see *Fig.*12).

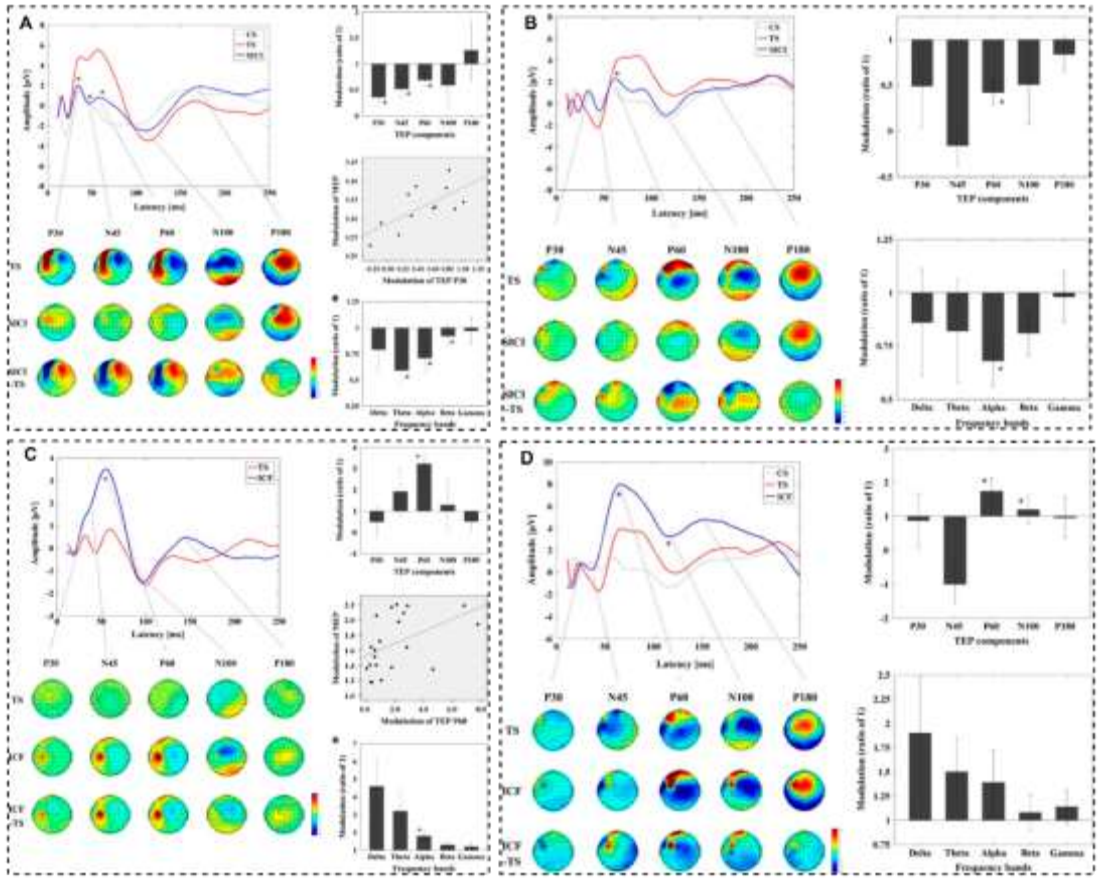

Fig. 10 Inhibitory/facilitatory influence of SICI and ICF on TEPs with TMS over M1 and DLPFC. (A) Inhibitory influence of SICI on TEPs with TMS over M1. (B) Inhibitory influence of SICI on TEPs with TMS over DLPFC. (C) Facilitatory influence of ICF on TEPs with TMS over M1. (D) Facilitatory influence of ICF on TEPs with TMS over DLPFC. Adapted from *R. F. H. Cash, et al.*, 2017[26].

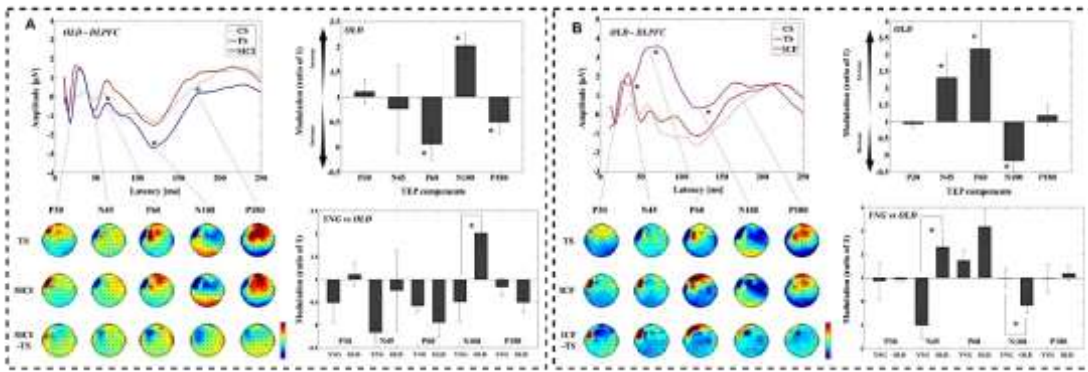

Fig. 11 Modulation of TEPs by the DLPFC-SICI & ICF paradigm following condition stimulus (CS), test stimulus (TS) in older (OLD) *vs.* young adults (YNG). (A) Modulation of TEPs by the DLPFC-SICI paradigm in older adults. (B) Modulation of TEPs by the DLPFC-ICF paradigm in older adults. Adapted from *Y. Noda, et al.*, 2017[27].

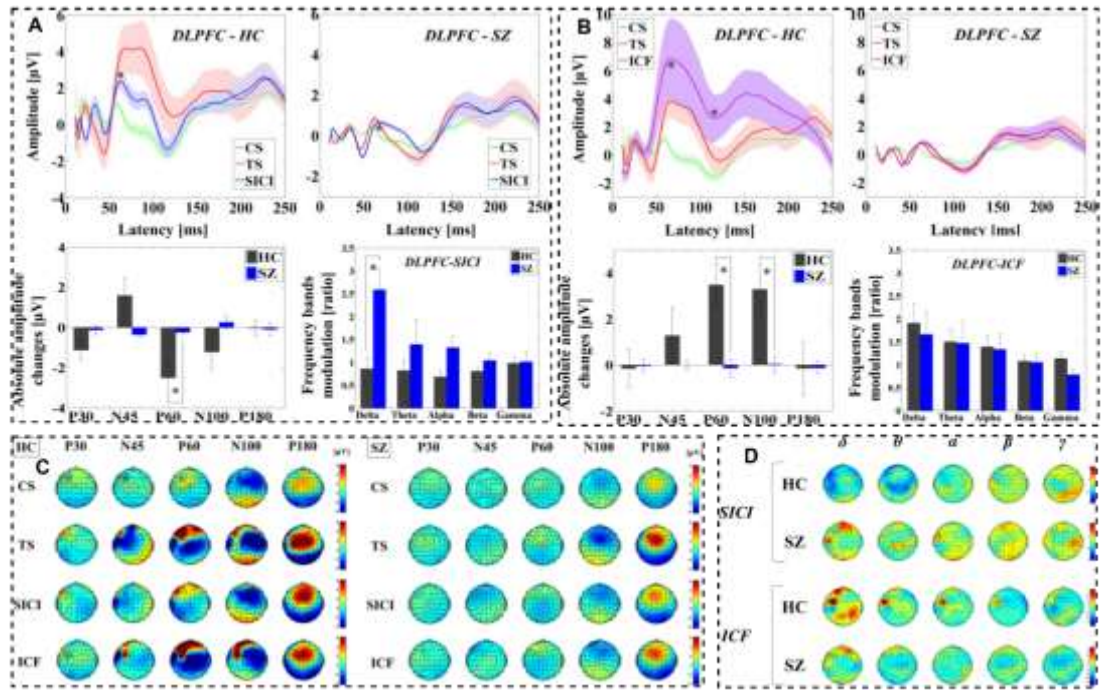

Fig. 12 SICI and ICF from the DLPFC in healthy control (HC) and schizophrenia (SC) following CS, test TS. (A) Modulation of TEPs by SICI paradigm administered TMS to DLPFC. (B) Modulation of TEPs by ICF paradigm administered TMS to DLPFC. (C) Topographical plots of paired pulse SICI and ICF paradigms. (D)Topographical distributions of frequency band modulations by SICI and ICF paradigms. Adapted from *Y. Noda, et al.*, 2017[28].

**Short latency afferent inhibition (SAI)**

SAI is obtained when applied to M1 with the combination of median nerve electrical stimulation and TMS precedes a TS at ISIs of 20-25 m leads to MEP suppression. SAI over M1 has been predominantly associated with cholinergic and $GABA_A$ergic circuits.
Assess correlates of SAI using TMS-EEG has indicated a decrease in the N100 component. If apply SAI protocol over M1, MEP amplitude reduce. A similar reduction of N100 amplitude accompanied by a P60 attenuation and a ERSP decrease in the beta band was found. Modulation of SAI associated with N100 component with MEP suppression, is an increase rather than a decrease was also observed (*Fig.*13-A).

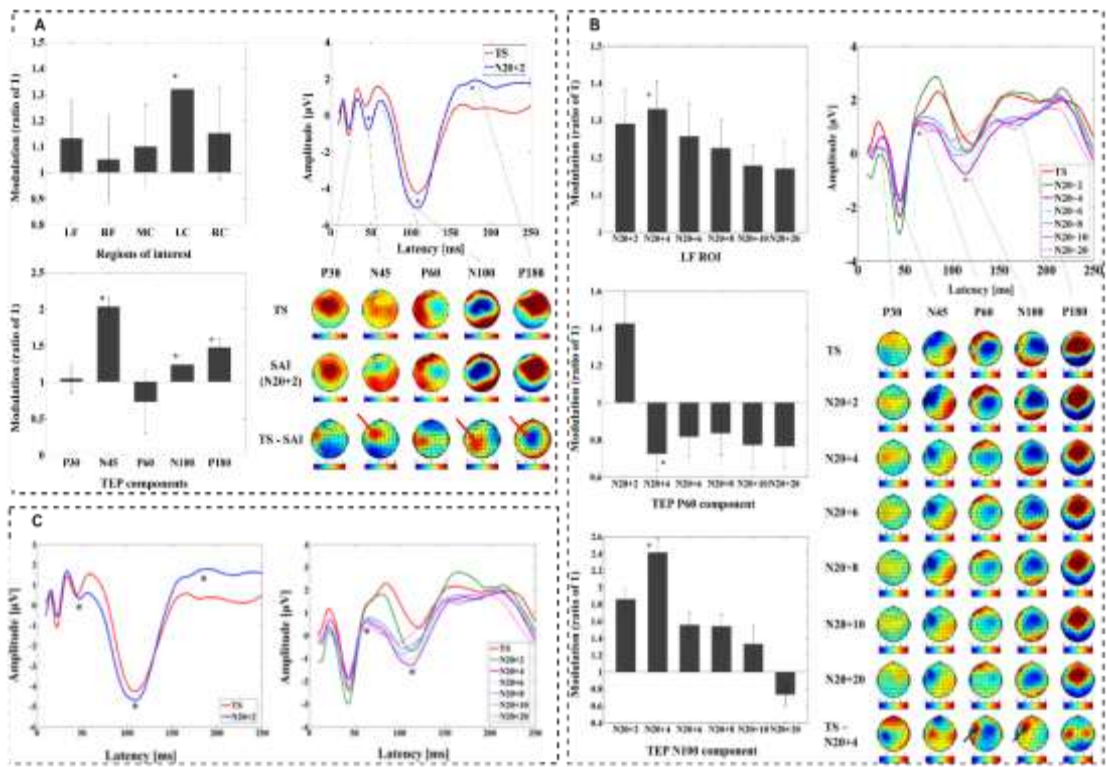

Fig. 13 TMS-EEG study of SAI in the M1 and DLPFC. (A) Modulation of cortical activity by SAI with TMS delivered to M1 (M1-SAI). (B) Modulation of cortical activity by SAI with TMS delivered to DLPFC (DLPFC-SAI).    (C) TEP traces in the SAI paradigm for stimulation of M1 (left) and DLPFC (right) stimulation without SSEP subtraction. Adapted and modified from *Y. Noda, R, et al.*, 2016[29].

*TMS-EEG brain stimulation protocols*

Different non-invasive transcranial brain stimulation (NTBS) protocols that modulate cortical circuits through plasticity-like effects include regular repetitive transcranial magnetic stimulation (rTMS), theta-burst stimulation (TBS), paired-associative stimulation (PAS) and transcranial direct current stimulation (tDCS).

Repetitive application of TMS pulses (rTMS) can assess neuroplasticity. Apply the repetitive pairing of TMS pulses to two brain regions, or to a sensory cortex with an appropriately timed peripheral sensory stimulus (paired associative stimulation, PAS) would induce spike-timing dependent plasticity. rTMS and PAS protocols of different stimulation frequency, pattern, location can enhance or suppress neural activity beyond the stimulation duration. Following active rTMS to the motor cortex, increases / decreases in MEP amplitudes in response to fixed intensity single-pulse TMS are thought to provide an index of long-term potentiation-like (LTP-like) /long-term depression-like (LTD-like) plasticity. Plasticity-inducing protocols can have behavioral effects and might be leveraged for therapeutic applications[20].

**Regular repetitive transcranial magnetic stimulation(rTMS)-EEG protocol**

rTMS entails the delivery of sequences or trains of magnetic pulses at diverse frequencies. When administered with an appropriate temporal pattern, duration, and intensity, these magnetic pulses in rTMS are anticipated to induce lasting changes in synaptic efficiency that persist beyond the stimulation period. Conventional rTMS paradigms typically employ suprathreshold pulses and extended sequences of stimuli lasting 10 to 25 min. When applied over motor regions, it is widely accepted low frequencies (≤1 Hz) predominantly results in suppression, while higher frequencies (5-20 Hz) tend to facilitate MEPs. One central hypothesis about the modulatory impact of rTMS in human cortex bear resemblance to long-term potentiation (LTP) and depression (LTD) observed in animal trials, which can be quantified *via* the response amplitude alterations to electrical extracellular electrodes stimulation (*Fig.*14).

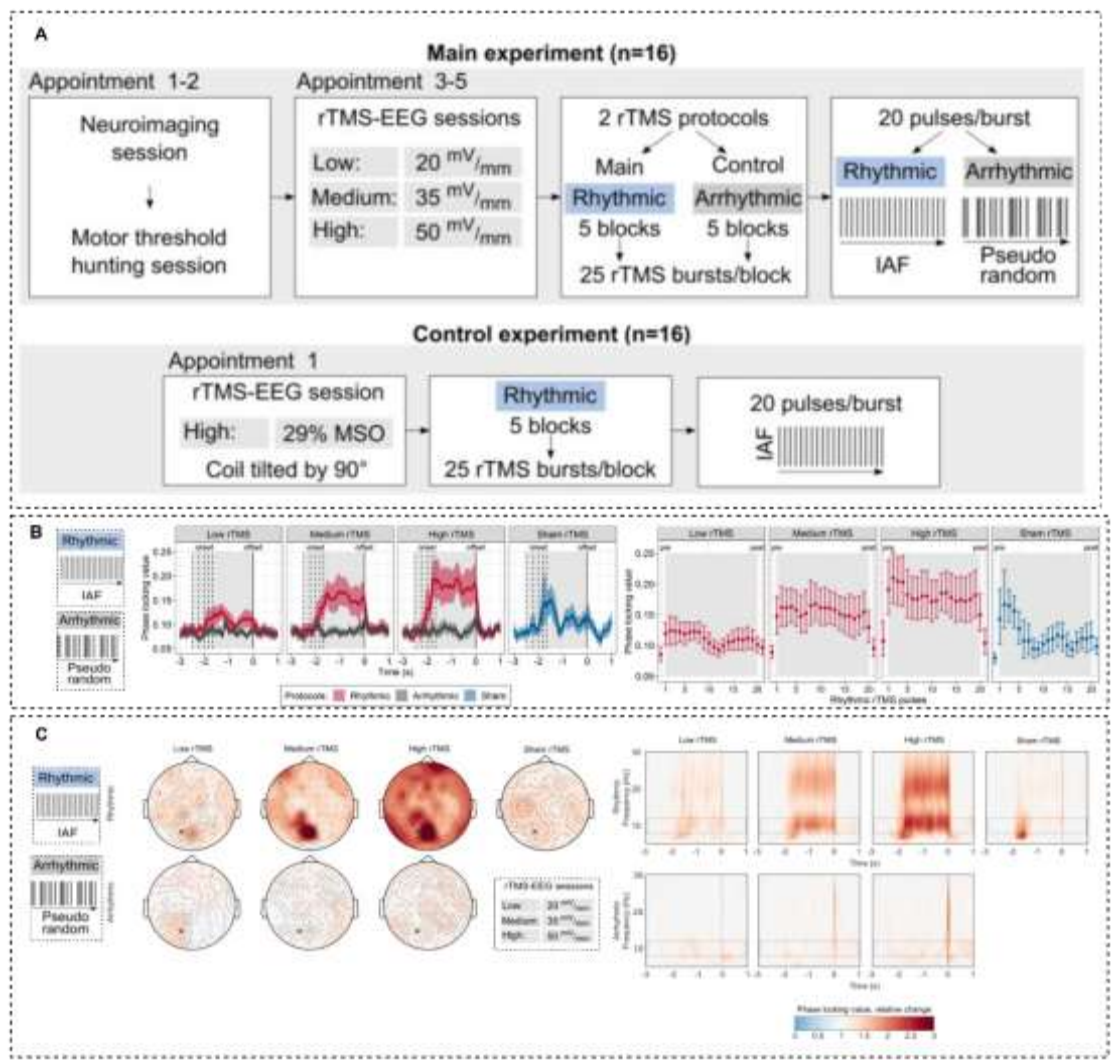

Fig. 14 rTMS-EEG experiment. (A) experiments setting. (B) Increased and sustained neural synchronization during rhythmic but not during arrhythmic or sham rTMS. (C) Rhythmic rTMS synchronized ongoing posterior $\alpha$ rhythms indicated by increased phase locking values. Adapted and modified from *E. Zmeykina, et al.*, 2020[30].

**Theta-burst stimulation (TBS)-EEG protocol**

TBS composed of three or more subthreshold stimuli at a high frequency within the theta range (30-50 Hz), which are then repeated at a lower carrier frequency (usually 5 Hz). $\theta$- frequency patterns have been shown to mimic the natural rhythms associated with synaptic plasticity mechanisms such as long-term potentiation (LTP) and long-term depression (LTD) according to animal studies. When applied intermittent TBS (iTBS), increases in corticospinal excitability are expected over motor regions. Inversely, when applied continuous TBS (cTBS) inhibition is observed[13]. To assess modulation of brain activity in response to TBS, investigating effects of active iTBS and cTBS protocols on resting state EEG in the same group of participants and compare to sham TBS could be explored (*Fig.*15).

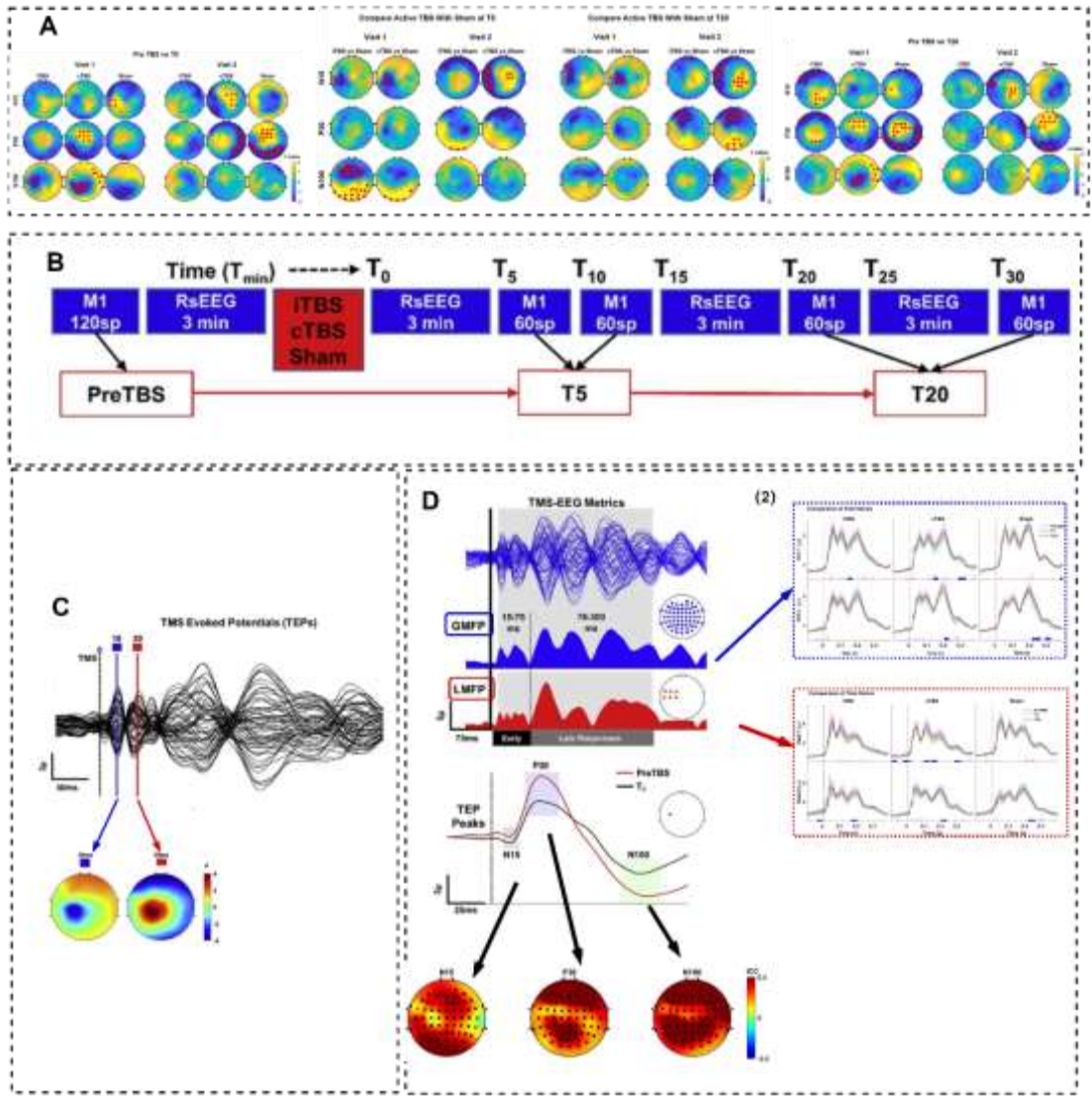

Fig. 15 TSB-EEG experiment. (A) TEP responses to different TBS at T5(left), T20(right) and sham control(middle). (B) TMS-EEG session protocol. (C) TMS evoked potentials from 63 channels (upper panel) with selected peaks (colored vertical lines) and topographical distribution of selected peaks of N15 & P30 (D) (1) Computation of GMFP & LMFP in left motor cortex within time-windows (early - late responses following TMS pulse), (2) Statistical comparison GMFP & LMFP responses at the millisecond level for both visit1 and visit2. (3) selected TEP peaks extracted from C3 electrode before (red line) and after (black line) iTBS at T20 in a representative subject (4) related topographical distribution of intra-class correlation coefficients of each TEP peaks across all visits at the electrode level. Adapted and modified from *R. A. Ozdemir, et al.,* 2021[31].

**Paired-associative stimulation (PAS)-EEG protocol**

PAS involves pairing a suprathreshold electrical stimulus applied to a peripheral nerve, usually the median nerve, with a suprathreshold TMS pulse applied to contralateral M1 using brief intervals known as ISIs. Modifying the ISI in PAS protocols effectively adjusts its impact, reflecting the principles observed in animal models of spike-timing-dependent plasticity (STDP). If a presynaptic input precedes postsynaptic excitation, synaptic transmission is facilitated, whereas, if postsynaptic excitation precedes a presynaptic input, transmission is inhibited. There is a general consensus that ISIs of 21.5-25 ms are facilitatory, aligning with a scenario where a pre-synaptic input preceding post-synaptic excitation, whilst shorter ISIs about 10 ms or so, are inhibitory.

PAS at an ISI of 25ms led to an enhancement of cortical excitability, as evidenced by increased GMFP and MEP amplitudes, not only in the hemisphere ipsilateral but also contralateral to the stimulation site. Conversely, PAS at an ISI of 10ms resulted in a decrease of GMFP and MEP amplitudes at the direct site of stimulation. Changes detected through TMS-EEG and TMS-EMG were characterized by a large inter-subject variability[13]. Combining TMS with EEG we aimed at investigating PAS effects and the connectivity modulation induced in humans (*Fig.*16).

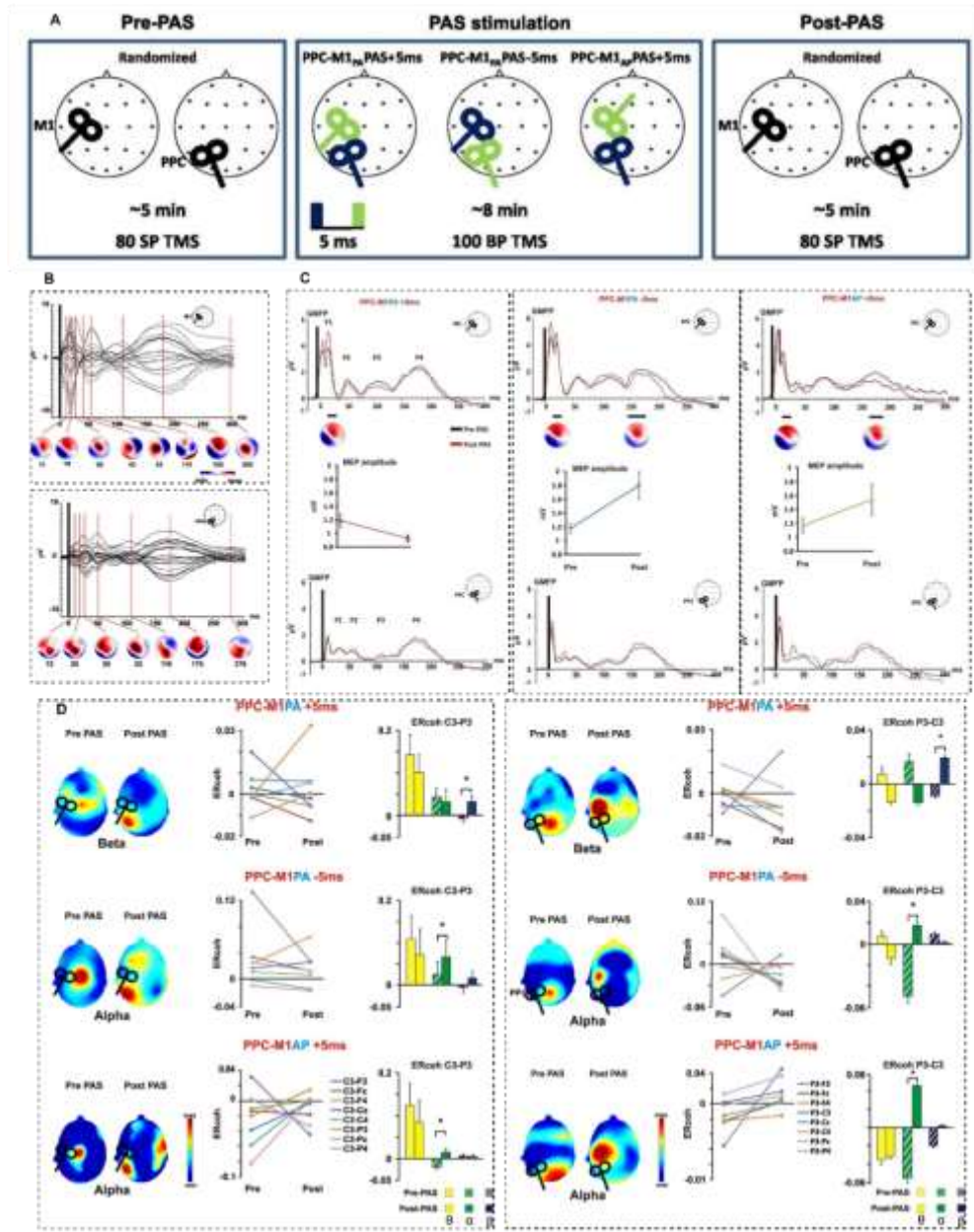

Fig. 16 PAS-EEG experiment. (A) Schematic representation of the experimental setting: eighty single pulse TMS were delivered before and after the administration of three different PAS protocols in a randomized order over M1 or PPC of the left dominant hemisphere. posterior-anterior (PA) to anterior-posterior (AP). (B) Grand-average of TEPs in a butterfly plot as a result of M1 (upper) and PPC (lower) stimulation with related distribution tomographs in specific time-domain. (C) Global cortical reactivity changes induced by PAS protocols of PPC-M1PA+5ms, PPC-M1PA-5ms and PPC-M1AP +5ms. (D) M1 (left) & PPC (right) event-related coherence changes induced by distinct PAS protocols. Adapted and modified from *D. Veniero, et al.*,2013[32].

## Conclusion

In summary, TMS-EEG methodology is wildly applied in neuroscience fundamental research and clinical implement due to the electrophysiological properties for studying neuron excitatory/inhibitory and neuronal population plasticity. Aspects of artifact correction, data analysis and clinical protocols of TMS-EEG are discussed for references considering to avoid pitfalls, though pros and cons. The future expectations would be combining more modern pharmaceutical and biological methods in exploring neuron conditions of all kinds in humans.

**Case Study of Analyzing Theta-Burst Stimulation to Frontal Cortex Based on Open Neuroscientific Data**

Hua Cheng

## Abstract

This study aims to leverage an openly available, high-quality EEG dataset to delve into the alterations in cortical activity. By applying Intermittent theta-burst stimulation (iTBS) and continuous theta-burst stimulation (cTBS) to the left dorsolateral prefrontal cortex (DLPFC) in healthy individuals, we observe changes in oscillatory patterns within the EEG data.
The dataset includes meticulously extracted resting-state EEG recordings, TMS-evoked potential data, and MRI scans. To process these data, we utilized Brainstorm, an open-source Matlab application, which facilitated noise reduction through independent component analysis and signal-space projection techniques. It allowed us to identify, visualize, and analyze TMS-evoked potentials (TEPs) and TMS-induced oscillations (TIOs). In addition, the study presents detailed plots of resting-state EEG power, local mean field power (LMFP), TMS-related spectral perturbation (TSRP), and inter-trial phase clustering (ITPC). Paired t-tests and cluster-based permutation tests have been performed for statistical analysis.
The wealth and quality of this dataset make it ideal for examining the neuromodulatory impact of TBS on the prefrontal cortex. Brainstorm's extensive feature set greatly supports the exploration of such neurological data. Future research directions could concentrate on conducting source localization analyses and comparative group studies.

## Data description

A comprehensive EEG dataset, openly accessible, offers data from both resting-state measurements and simultaneous single-pulse TMS-EEG sessions. This dataset facilitates an investigation into the alterations of cortical activity resulting from the application of Intermittent theta-burst stimulation (iTBS) and continuous theta-burst stimulation (cTBS) to the left dorsolateral prefrontal cortex (DLPFC) in individuals with no health issues. TBS intervention modifies oscillatory patterns within the specific frequency bands inherent to this type of intervention (*i.e.*, 5 Hz and 50 Hz).

The EEG data were acquired using a *Refa* 2048 *Hz* EEG system and an appropriately sized 64-channel 10–20 EEG cap as determined by head circumference, with sintered, interrupted disk, Ag-AgCl TMS-compatible electrodes. The position of the EEG cap was confirmed by matching the ***Cz*** electrode with the intersection of the participants' nasion-inion and tragus-tragus axes. Electrodes were grounded to ***Fpz***, and EEG signals were measured against a common average reference. To reduce scalp impedance, participants were instructed to wash their hair before attending each experiment session. Secondly, prior to cap placement, the participant's scalp was cleaned with alcohol swabs. Lastly, an electro-conductive gel and blunted needles were used to lightly abrade the scalp to limit impedances to less than 50 $k\Omega$, which is well below 1% of the input impedance ($100 M\Omega$) of the EEG amplifier[33].

TBS and single-pulse TMS were delivered using a *MagPro® X100*. with a 65 mm diameter *Cool-B65 figure-8* stimulation coil. The coil was positioned tangentially to the scalp over the **F3** electrode in order to target the DLPFC. A 5-mm customised 3D-printed spacer was placed between the coil and the scalp at all times to avoid contact with electrodes to minimise post-pulse artefacts, electrode movement, and bone-conducted auditory input. The coil was oriented at a 45-degree angle relative to the parasagittal plane, and the TMS pulse was delivered using a biphasic waveform. A hard foam headrest connected to a mechanical arm was positioned on the contralateral temporal region of the stimulation site to ensure minimal participant movement[33]. Open dataset② files are stored in *. mat* format and contain raw EEG data. The data in each file includes 68 labelled signals (64 EEG channels, ECG, HEOG, VEOG, and a Trigger channel to mark events). Experiment blocks are labelled using the naming structure, as below *Tab*.1.

Tab. 1 Naming structure of EEG task blocks for each session[33].

| Steps | Experiment Block Sequence | Filename Format | Blocks type | Time courses |
|---|---|---|---|---|
| 1 | Eyes-open resting-state EEG | pre-rest_run-01 | RS-EEG | 5 |
| 2 | Single-pulse TMS-EEG | pre-tep_run-01 | TMS-EEG (pre) | 10 |
| 3 | Eyes-open resting-state EEG | pre-rest_run-02 | RS-EEG | |
| 4 | cTBS/iTBS/Sham | tbs | Condition intervention | 5 |
| 5 | Single-pulse TMS-EEG | post-tep_run-01 | TMS-EEG (T2) | 15 |
| 6 | Eyes-open resting-state EEG | post-rest_run-01 | RS-EEG | |
| 7 | Single-pulse TMS-EEG | post-tep_run-02 | TMS-EEG (T15) | 15 |
| 8 | Eyes-open resting-state EEG | post-rest_run-02 | RS-EEG | |
| 9 | Single-pulse TMS-EEG | post-tep_run-03 | TMS-EEG (T30) | 10 |
| 10 | Eyes-open resting-state EEG | post-rest_run-03 | RS-EEG | |

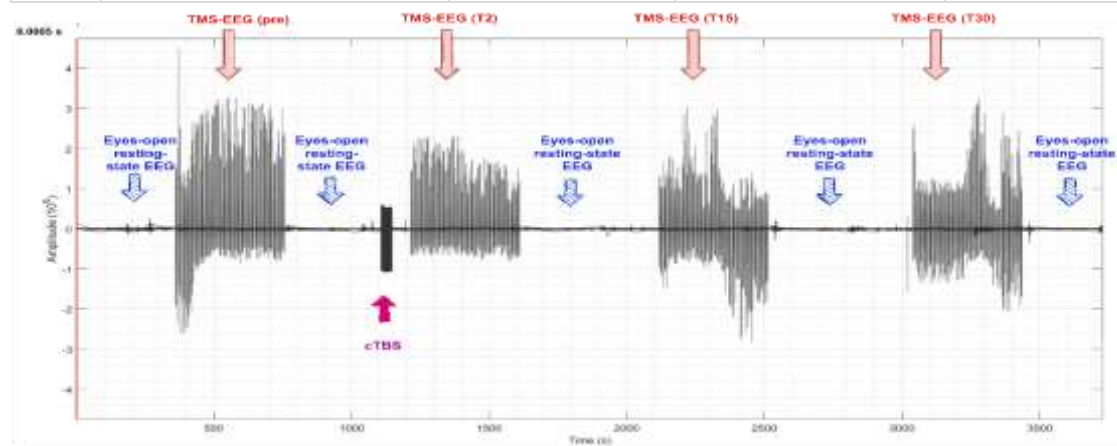

Fig. 17 cTBS protocol.

---

② The raw data files and code can be accessed via the FigShare open access repository service (https://doi.org/10.25452/figshare.plus.c.5910329). Url: https://doi.org/10.25452/figshare.plus.c.5910329.v1.

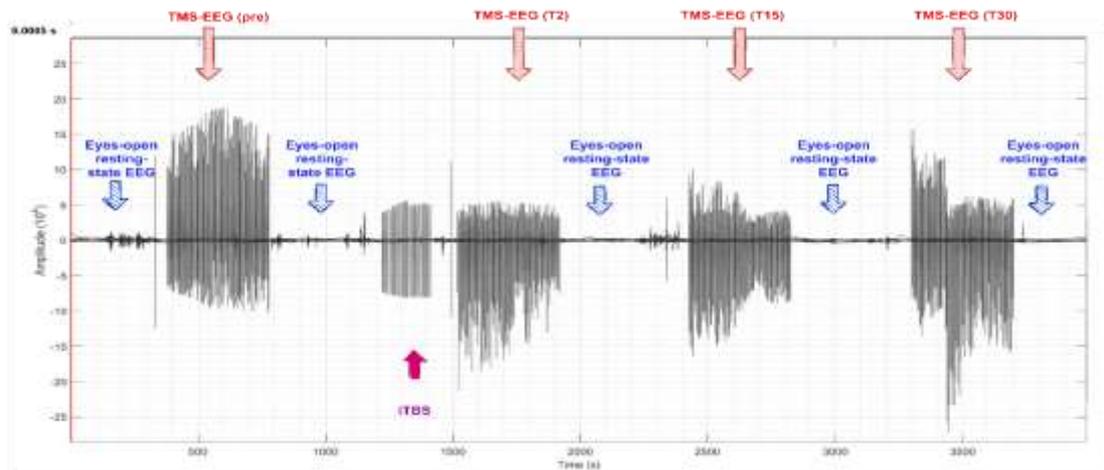

Fig. 18 iTBS protocol.

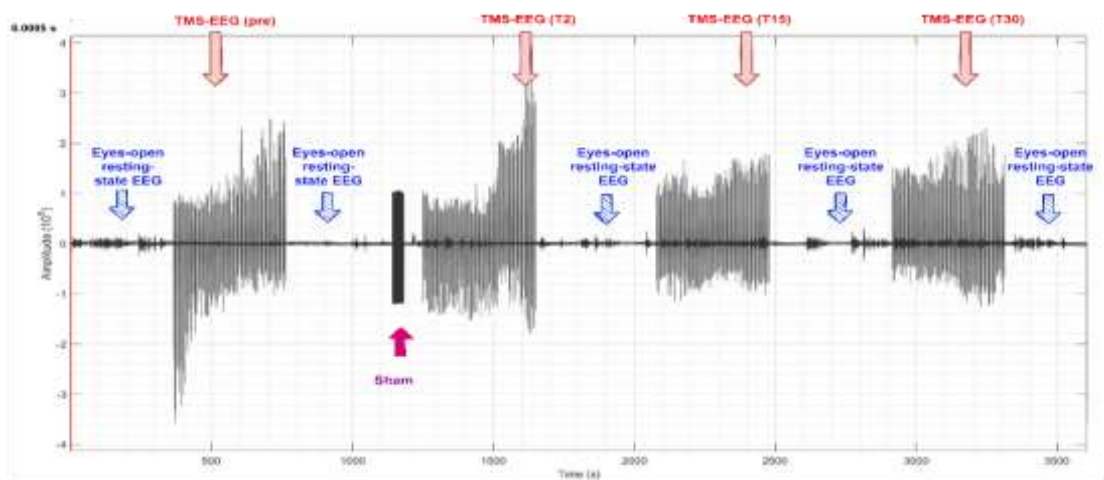

Fig. 19 Sham stimulation protocol.

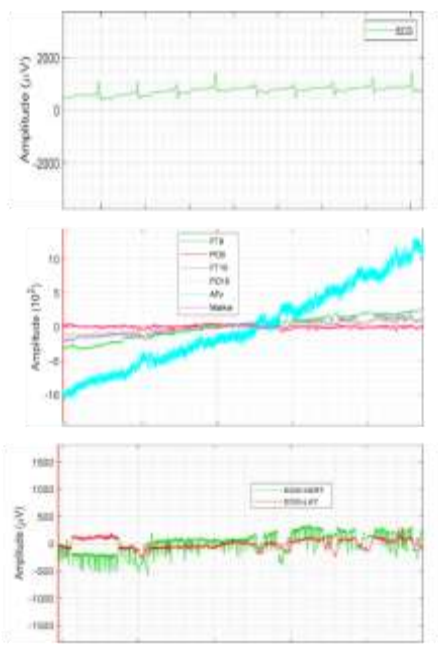

Fig. 20 ECG channel(top), EEG_NO_LOC channels(middle) and EOG EOG-VERT&EOG-LAT

channels(bottom).

Additional files include: a generic description of the metadata (*dataset_description.csv*), participant demographic and stimulation parameter details (*participants.csv*), metadata of the experiment tasks and EEG recording system (*eeg.csv*), a list of all EEG channels (*channels.csv*), neuronavigated coordinates (*electrodes.csv*), source data is provided in the MATLAB file format (*.mat*).

## *Resting-state EEG (RS-EEG) data*

1) Eyes-open RS-EEG data were down-sampled to 512 Hz, baseline-corrected (demeaned) and detrended.
2) Then remove electrical line noise by using a second-order bandpass filter (0.1–70 *Hz*) and a notch filter at 50 *Hz*.
3) RS-EEG data were epoched in 1-s intervals. Reject epochs by using an automated algorithm in which epochs with data ranges greater than 3 standard deviations (SD) or absolute maxima greater than 12 SD of other epochs.
4) Reject remaining noisy epochs with a visual inspection.
5) Remove components containing eye blinks and muscle artefacts with a single round of independent component analysis (ICA).
6) Re-referenced EEG data to the common average reference.
7) Power spectral densities (PSD) were calculated using 180 s of data. Log-normalised power spectral density values ($\mu V^2/Hz$) were estimated for each EEG electrode over a range of 1–70 *Hz* using the fast Fourier transform (FFT) with 2-s sliding Hamming windows with 50% overlap.

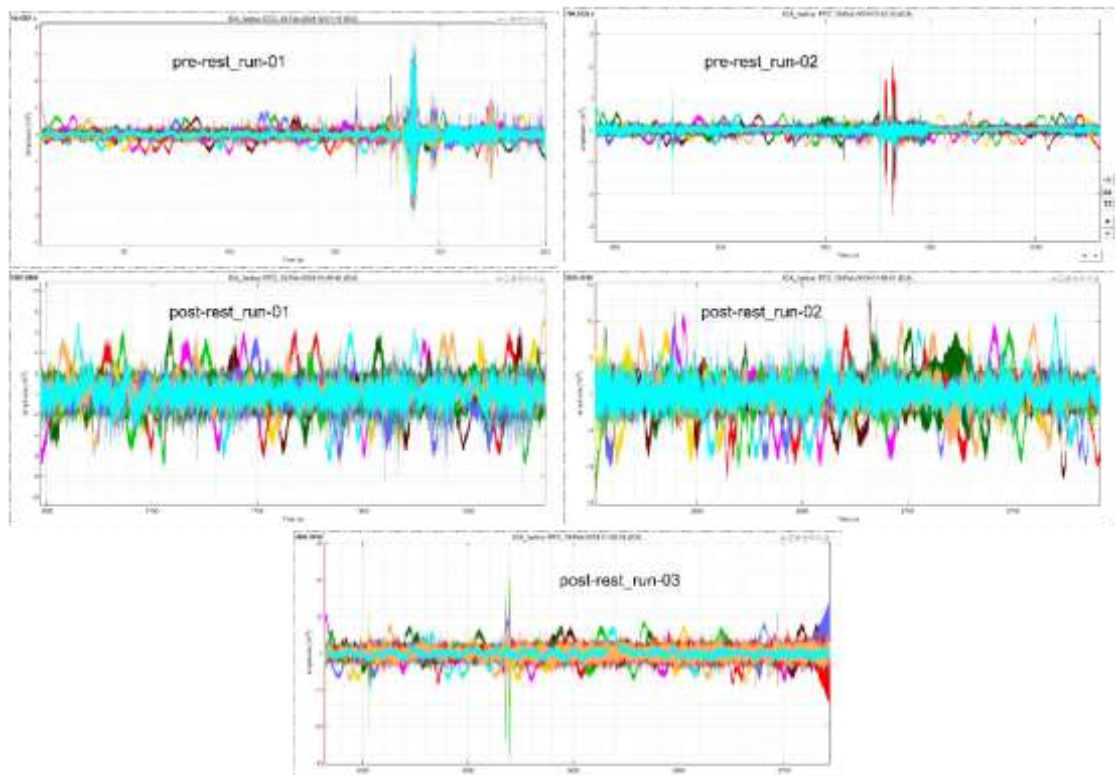

Fig. 21 Eyes-open resting-state EEG.

## TMS-evoked potentials (TEPs) data

A double-blinded crossover design using five different occasions, includes sessions of 2 iTBS, 2 cTBS and 1 sham session. A burst of 3 pulses at 50Hz with 200 ms between bursts in a total of 600 pulses. The iTBS protocol involved a 2 s train of TBS repeated every 10 s for a total of 190 s (600 pulses), and cTBS consisted of a 40 s train of uninterrupted TBS (600 pulses). Sham TBS consisted of an inactive coil positioned on the head (at the same position as the active conditions) and a second active coil positioned 20 cm from the back of the head, facing away from it, with an increased stimulation output of 20% to compensate for the attenuation of the sound due to the additional distance from the ear.

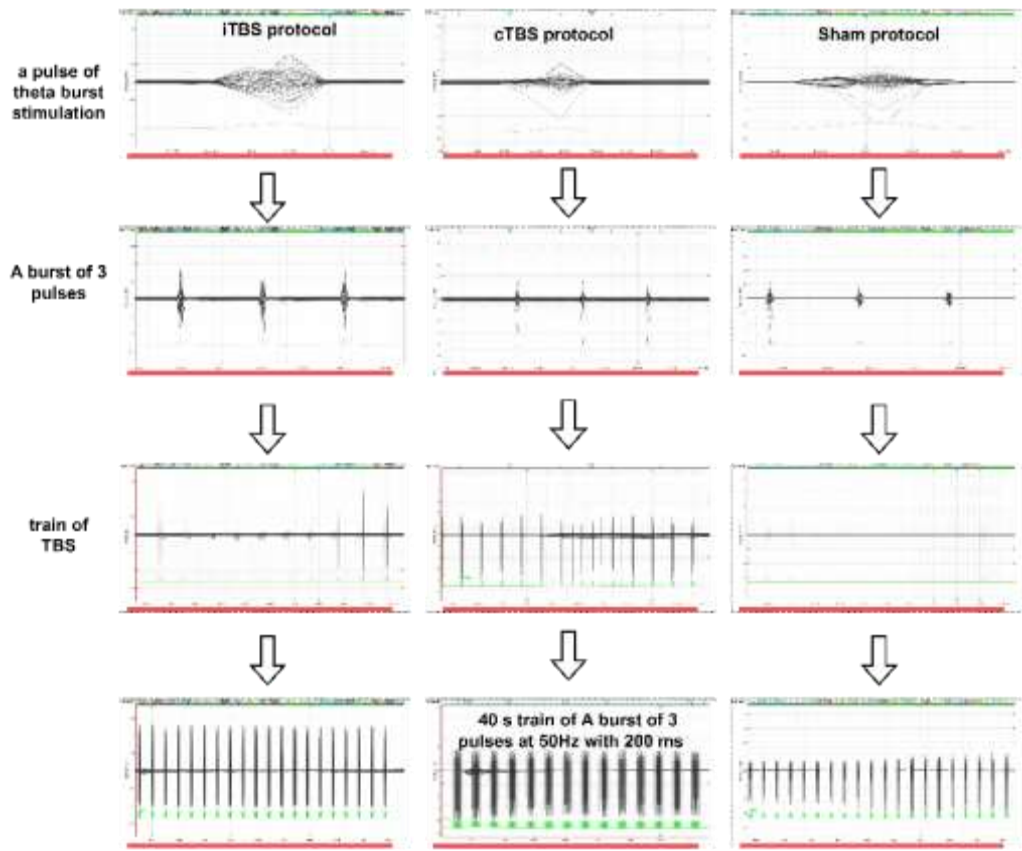

Fig. 22 TBS Experimental protocols.

Record the EEG responses to 100 single TMS pulses(sTMS) before each TBS condition (baseline block) and at 2-, 15- and 30-min post-TBS (T2, T15 and T30, respectively)[34].

1) TEPs were epoched around the TMS pulse (−1000 to 1000 ms).
2) Electrodes in which the TMS artefact exceeded the maximum absolute value of the range of the amplifier ($10^7 \mu V$) were removed and linearly interpolated from neighboring channels.
3) EEG traces were detrended and baseline-corrected relative to pre-TMS data (−500 to −50 ms).
4) Line noise (50Hz) was removed using linear regression by fitting and subtracting a sine wave from the EEG.
5) Data between −5 and 10 ms around the TMS pulse were removed.
6) An initial round of ICA was performed using the TESA *compselect* function to eliminate components containing eye blinks.

7) TMS-muscle and decay artefacts were removed by fitting a power law to the most negative and positive EEG signal deflections caused by the TMS artefact to obtain regression fit parameters, and then removing the artefact from the data by subtraction.
8) EEG data from before and after the TMS stimulus were filtered separately using a bandpass filter (1–90 Hz).
9) A second round of ICA was performed to remove this as well as components associated with blinks, eye movement, persistent muscle activity, decay artefacts and electrode noise.

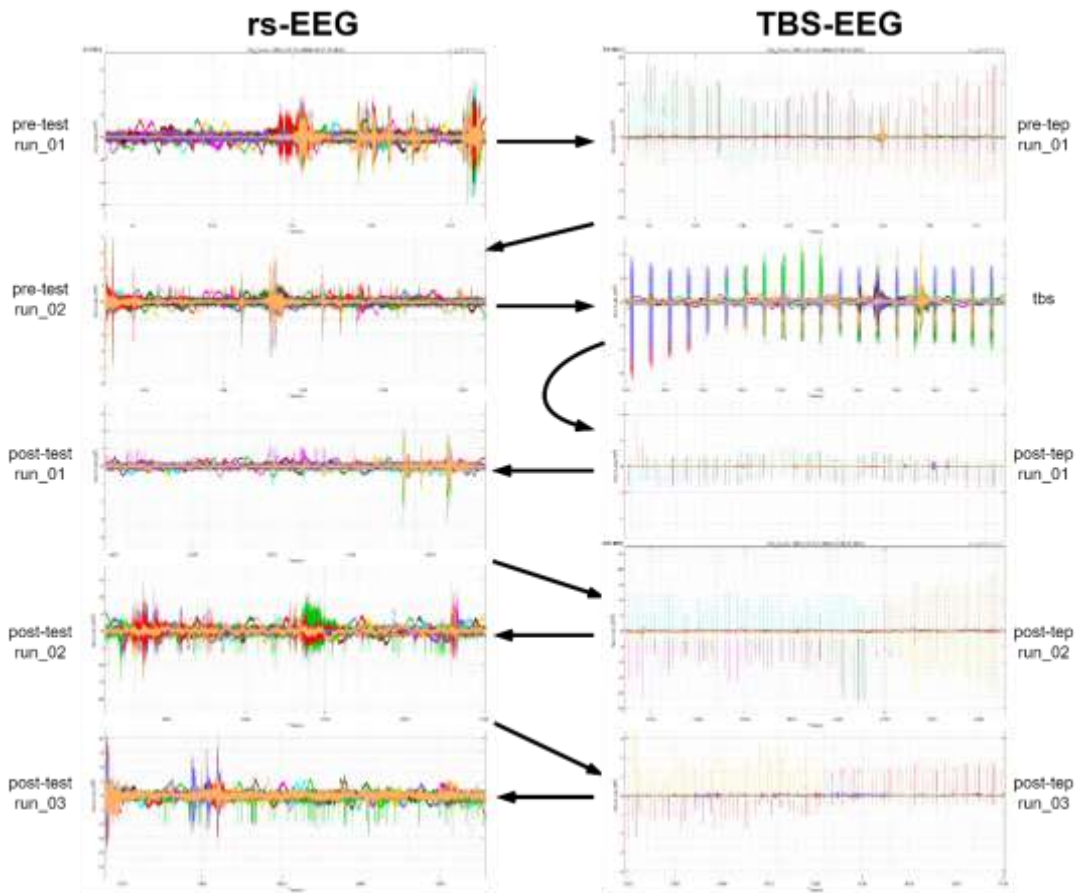

Fig. 23 Resting state EEG (left column) and Transcranial Magnetic Stimulation evoked EEG (right column) in different epochs. Black arrows represent the time flow.

*MRI scan data*

The MRI dataset can be used to assess how individual differences in brain morphology, including white matter fibre bundle size, grey matter volume, and whole brain volume, may affect the TMS-evoked potential. The MRI scans were obtained using a *Philips Achieva 3 T (TX) - DS MRI* scanner based at *Neuroscience Research Australia (NeuRA), Sydney, Australia*. A single scan was obtained for each participant at baseline prior to the start of the TMS sessions[33].

1) All MRI scans have been de-identified and anonymised using the **Fieldtrip** *ft_defacevolume* and *ft_anonymizedata* functions.
2) For all participants, *T1-weighted sequences* (TR=5.7ms, TE=2.6ms, FOV=250 ×250×190 mm, voxel size=1×1×1mm, matrix 250×250, Flip angle 8°, 190 sagittal plane slices) were used to acquire structural MR images covering the whole brain.
3) In the same session, *high-resolution DTI* (TR=13737ms, TE=59ms, FOV=240 ×240×120 mm, voxel size=2×2×2mm, matrix 120×120, Flip angle 90°, 30 transverse plane slices) was also acquired.

Tab. 2 Naming structure of MRI scans for each session[33].

| Experiment Block Sequence | Filename Format | type |
|---|---|---|
| **3D Ultrashort Echo Time sequence** | 3DUTESkull2mmiso | MRI scan |
| **Susceptibility weighted imaging** | sWIP3DUTESkull2mmiso | MRI scan |
| **Diffusion Tensor Imaging** | DTI | DTI scan |
| **T1-weighted image** | T1075TFESag | MRI scan |

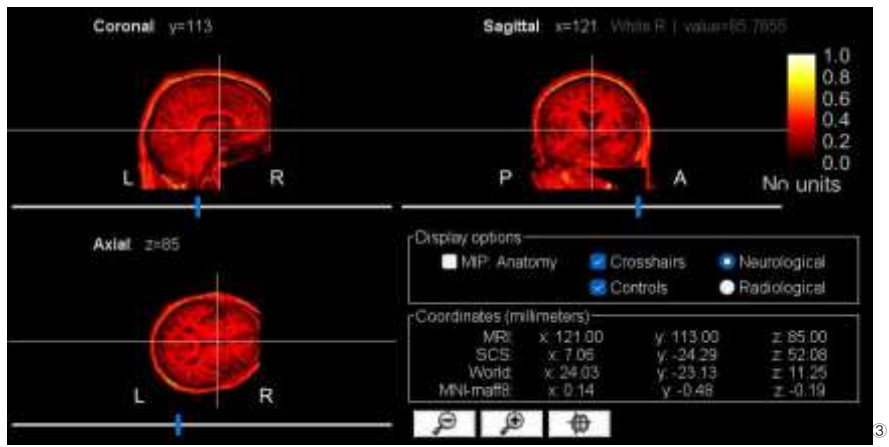

Fig. 24 MRI scan (from subject01).

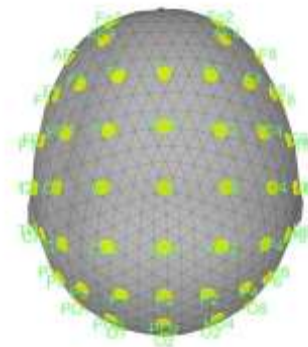

Fig. 25 EEG head sensors display.

## Data process

Initially, upon importing the data, the ICBM152 2023b[④] template was incorporated as the default anatomical reference after warping. Subsequently, the process involved identifying physiological events such as heartbeats and eye blinks. Following this step, a series of filtering procedures were implemented for preprocessing purposes, aimed at cleaning and excluding unwanted signals from the ECG (electrocardiogram) and EOG (electrooculogram) channels.

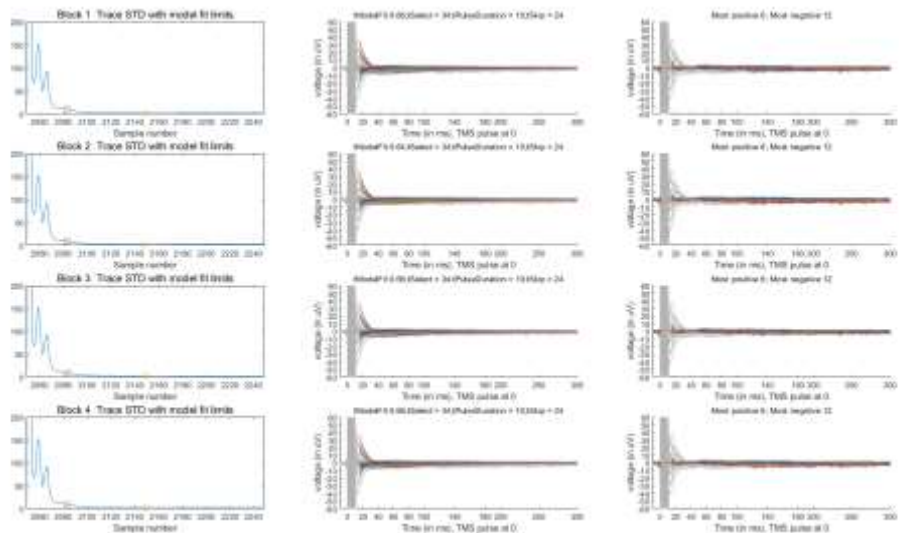

Fig. 26 Regression removal of TMS artefact.

### *Independent component analysis (ICA)*

When conventional frequency filters fail to eliminate transient artifacts or those that spectrally overlap with the targeted brain signals, Independent Component Analysis (ICA) can be employed. This method discerns unique spatial patterns associated with artifacts and subsequently separates them from the EEG recordings. The key aspect of ICA is that it isolates components that are temporally independent. Alternatively, Signal-Space Projection (SSP) is another technique that can be used to correct for such artifacts, providing additional strategies to address these challenges in EEG data preprocessing.

---

[④] Download - Brainstorm (usc.edu).

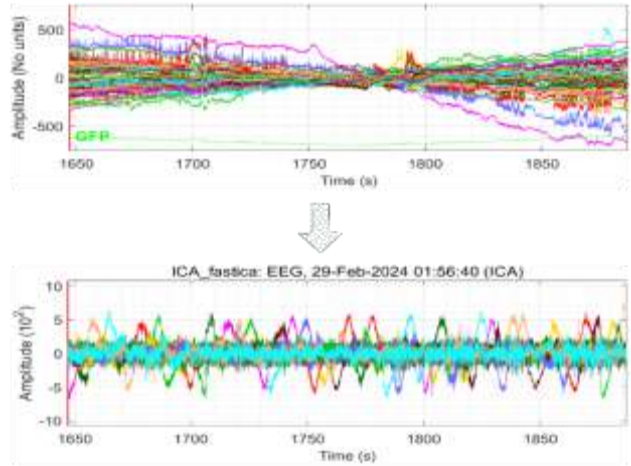

Fig. 27 ICA decomposition.

*Signal-space projection (SSP)/Principal components analysis (PCA)*

Signal-Space Projection (SSP) serves to pinpoint the characteristic sensor distributions linked to particular artifacts and generates spatial filters to effectively subtract the influence of these patterns from the recorded data. Implementing a Principal Components Analysis (PCA) on a concatenated set of artifacts aids in decomposing the distinct spatial elements present. In the case of removing heartbeats and eye blinks, SSP provides a streamlined solution that represents a subset of a broader, more generalized artifact correction procedure. The percentage (%) indicates the amount of signal ($Si$) that was captured by the component during the decomposition: $\left(\% = \frac{Si}{\sum Si}\right)$ .

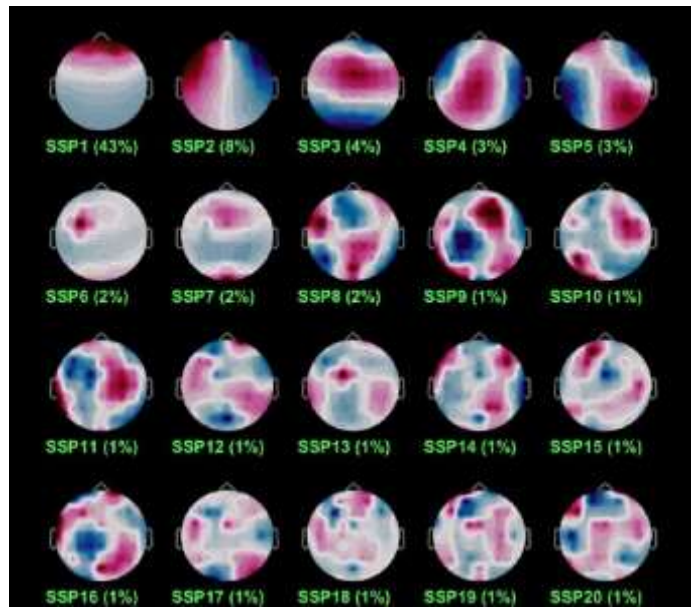

Fig. 28 Distribution topology plotting change with SSP/PCA decomposition percentage.

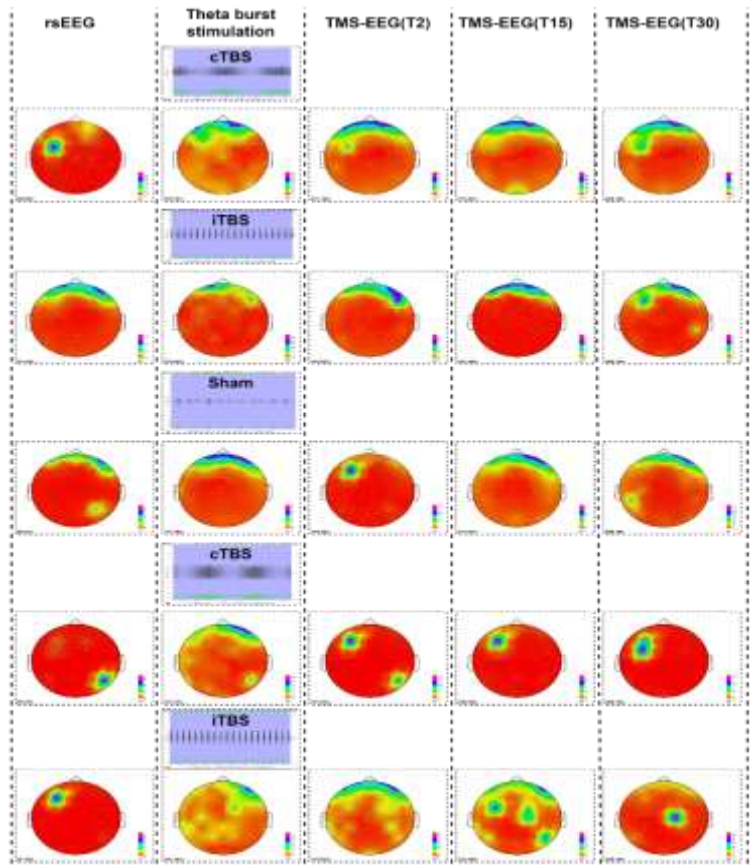

Fig. 29 Average distributions of subject 01 in difference condition.

## TMS-evoked EEG potentials (TEPs)

TMS-evoked EEG potentials (TEPs) refer to the EEG responses elicited by Transcranial Magnetic Stimulation (TMS), which are obtained through the temporal averaging of EEG signals in the time domain[35].

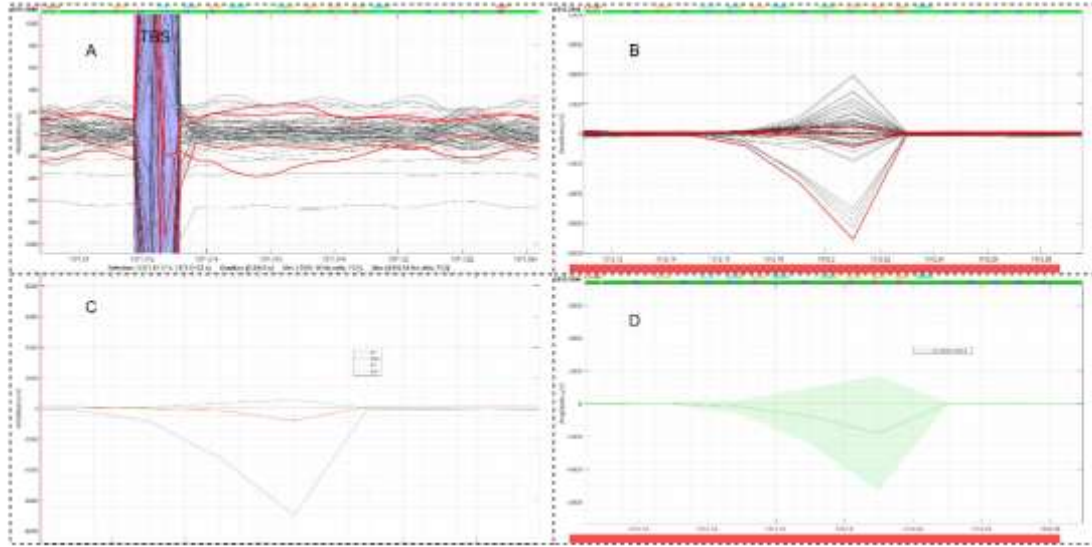

Fig. 30 TEPs plotting in time series. (A) Butterfly plot from all electrodes of one pulse of TBS and evoked potentials. (B)one pulse of a TBS (red lines represent channel of F1, F3, FC1, FC3). (C)one pulse of TBS in channels of interested (F1, F3, FC1, FC3). (D)Mean+Sdt of four electrodes (F3, FC3, F1, FC1).

## TMS-induced oscillations

The time-frequency analysis of TMS-EEG data unveils TMS-induced oscillations that encode stimulus-phase-locked information, initially marked by a surge in δ(delta), θ(theta), α(alpha), and β (beta) band power, succeeded by suppression or de-synchronization in α and β bands, and ultimately culminating in an augmentation of β band power.

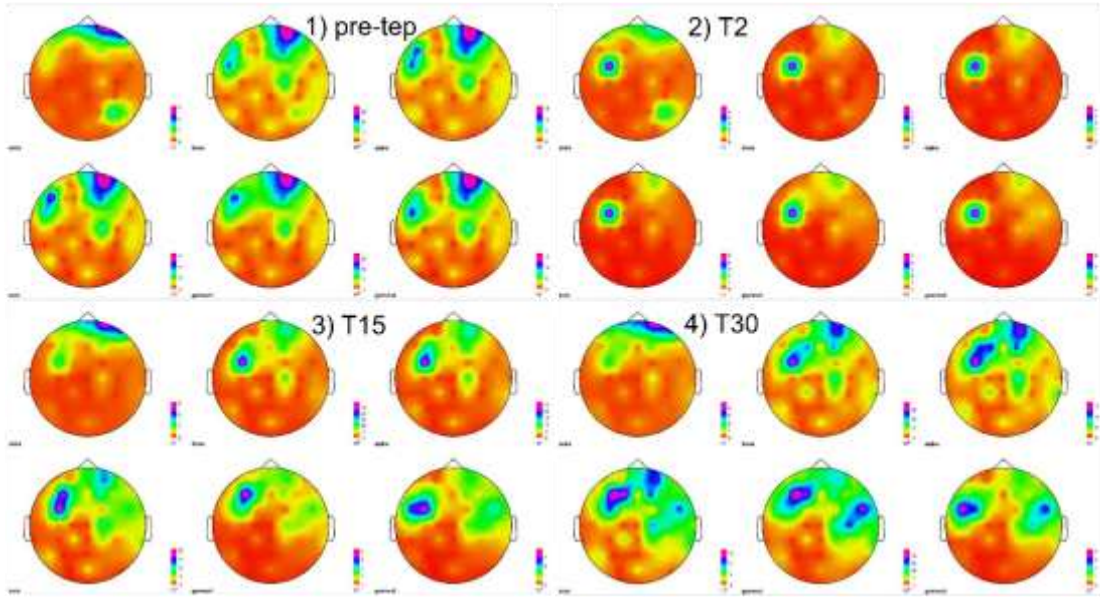

Fig. 31  Topology distribution plots of cortical status of $\delta$(2-4Hz),$\theta$(5-7Hz),$\alpha$(8-12Hz),$\beta$(15-29Hz),$\gamma$1(30-59Hz),$\gamma$2(60-90Hz) in the time point of (1) pre-TMS-EEG, (3)T2, (4) T15, (5)T30 in Sham condition.

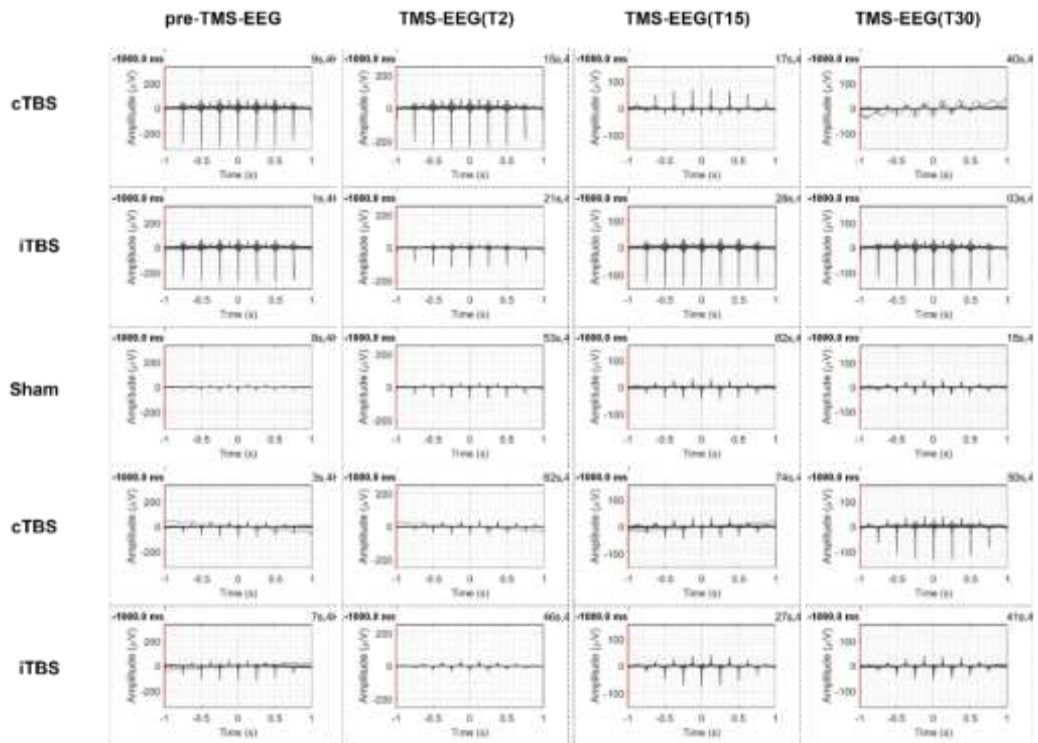

Fig. 32  Time-frequency decomposition visualizations of subject01. Assess the phase-amplitude coupling in different conditions. The nesting frequency (low) is set to 4Hz.

## *Resting state EEG power，LMFP，TRSP and ITPC*

Resting EEG measures combining TMS-EEG might represent a more thorough reflection of cortical excitability. The local mean field power (LMFP) as the square root of squared TEPs averaged across the four channels of interest. TMS-related spectral perturbation (TRSP) was evaluated locally by averaging the values obtained by the electrodes surrounding the stimulation site (FC3, FC1, F3, F1). The amount of TMS-related spectral perturbation (TRSP) was computed as[36]:

$$TRSP(f,t) = \frac{1}{n}\sum_{k=1}^{n}|F_k(f,t)|^2 \quad (1)$$

Inter-trial phase clustering (ITPC) was computed according to[36]:

$$ITPC(f,t) = \frac{1}{n}\sum_{k=1}^{n}\frac{F_k(f,t)}{|F_k(f,t)|} \quad (2)$$

In Equ (1) & (2), according to Delorme and Makeig[37], for $n$ trials, the spectral power or amplitude estimates $P$ and $F$ were computed at trial $k$, at frequency $f$ and time $t$.

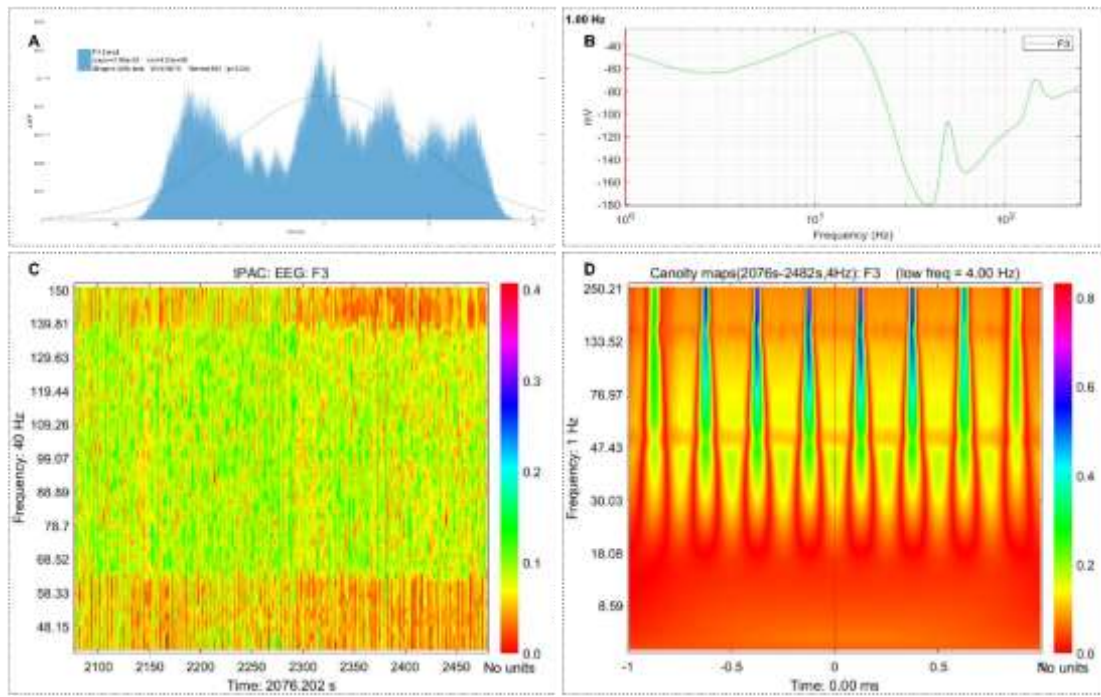

Fig. 33 Example of F3 from subject-01_session-03_post-tep_run-02. (A) LMFP; (B)resting EEG power; (C) TRSP; (D) ITPC.

## Statistical analysis

### *Paired t tests*

Reliability was tested in repetitive tests within same conditions. Differences of means in two visits of the same subject ware made.

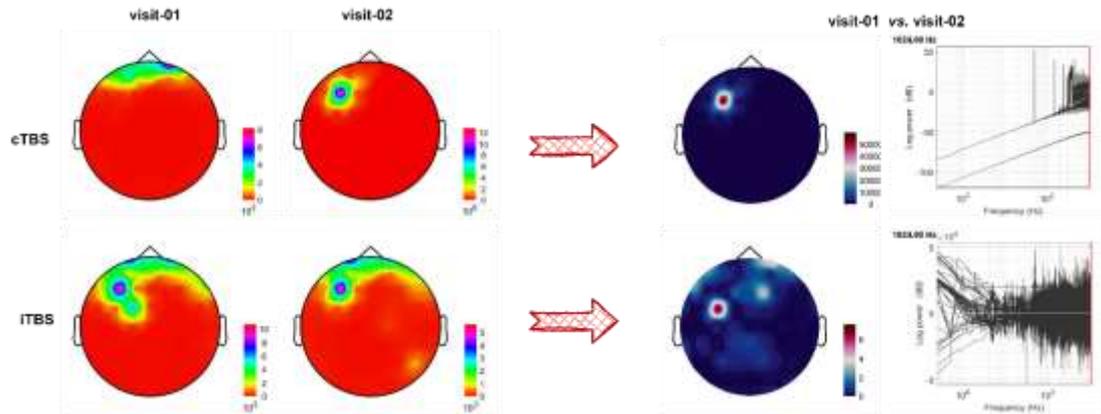

Fig. 34  Comparation between grand average of topological distributions of subjects in visit_01 and visit_02(mean=-4.66x10$^5$, std=3.33x10$^7$, P<0.05).

No significant differences were detected between pre- *vs.* post- TEP data within the same TBS condition(P<0.05).

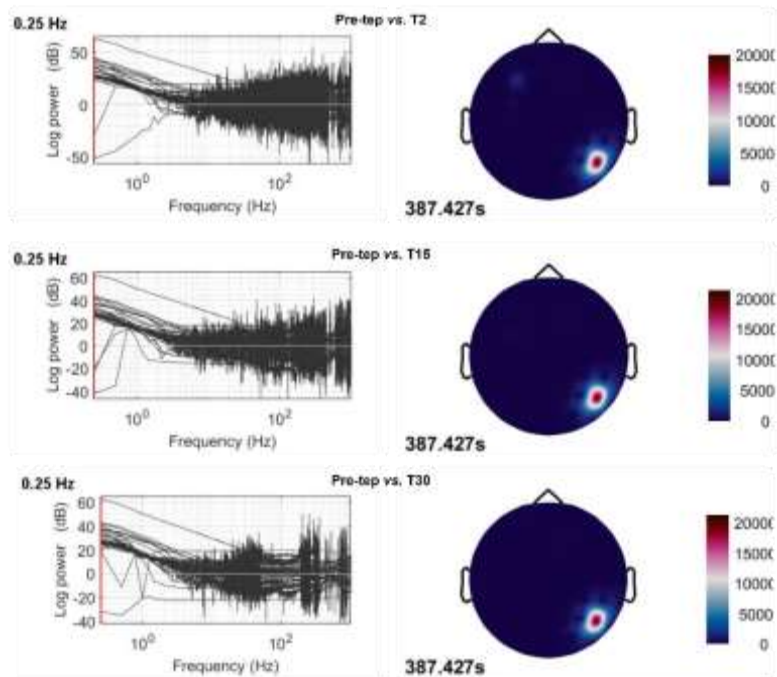

Fig. 35 pre-vs. post- TEP data within the same cTBS condition of sub01. (upper row) pre-TEP *Vs*. T2-TMS-EEG data. (middle row) pre-TEP *Vs*. T15-TMS-EEG data. (bottom row) pre-TEP *Vs*. T30-TMS-EEG data.

*Cluster-based permutation tests*

Statistics between conditions (cTBS *vs.* iTBS *vs.* Sham) for each electrode (FC1, FC3, F1, F3) in a selected a region of interest around the stimulation site and the corresponding contralateral site and each frequency of interests: delta (2~4 Hz), theta (4~7 Hz), alpha (8~12 Hz) beta (13~30 Hz), gamma (30~45 Hz) and gamma2 (46~90 Hz).

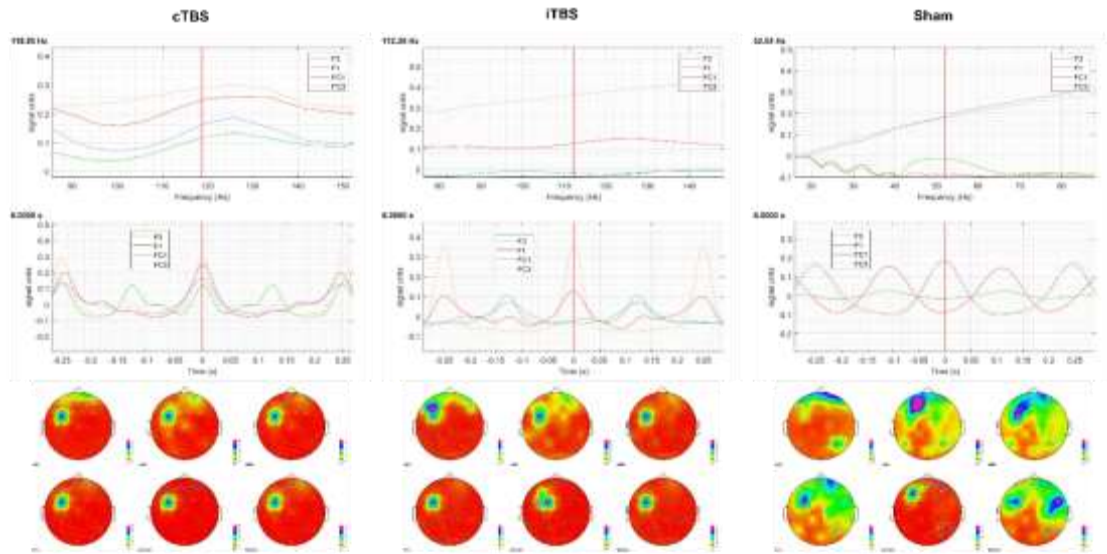

Fig. 36　Time-frequency-channels triplet compared. (upper row) power spectrum density of selected channels; (middle row) time series of selected channels;(bottom row) specific frequency band topological distribution of selected channels.

## Conclusion

Assessing and monitoring the influence of pharmacological agents and non-invasive brain stimulation techniques, such as TMS, on brain activity and cortical networks via TEPs data is a widely adopted method to gauge the functional state of both healthy and diseased individuals. In this study, a high-quality, openly accessible TMS-EEG dataset was employed to investigate the neuromodulatory effects of theta burst stimulation on the prefrontal cortex. The Brainstorm and Fieldtrip toolkits integrated into MATLAB R2022b leveraged for data preprocessing, analysis, and visualization. While certain source EEG data were managed to extract and analyze to reveal dynamic changes following diverse events, there is yet exploited potential to extract and reconstruct more data with the obstacles overcome in the advancing of computational resources.

## References

1　Hallett, M. Transcranial magnetic stimulation and the human brain. *Nature* **406**, 147-150, doi:10.1038/35018000 (2000).

2　Hernandez-Pavon, J. C. *et al.* TMS combined with EEG: Recommendations and open issues for data collection and analysis. *Brain Stimulation* **16**, 567-593, doi:https://doi.org/10.1016/j.brs.2023.02.009 (2023).

3　Rogasch, N. C., Daskalakis, Z. J. & Fitzgerald, P. B. Cortical inhibition, excitation, and connectivity in schizophrenia: a review of insights from transcranial magnetic stimulation. *Schizophr Bull* **40**, 685-696, doi:10.1093/schbul/sbt078 (2014).

4　Lioumis, P., Kičić, D., Savolainen, P., Mäkelä, J. P. & Kähkönen, S. Reproducibility of TMS-Evoked EEG responses. *Hum Brain Mapp* **30**, 1387-1396, doi:10.1002/hbm.20608 (2009).

5　Kallioniemi, E., Saari, J., Ferreri, F. & Määttä, S. TMS-EEG responses across the lifespan: Measurement, methods for characterisation and identified responses. *Journal of Neuroscience Methods* **366**, 109430, doi:https://doi.org/10.1016/j.jneumeth.2021.109430 (2022).

6       Ferreri, F. *et al.* Human brain connectivity during single and paired pulse transcranial magnetic stimulation. *NeuroImage* **54**, 90-102, doi:https://doi.org/10.1016/j.neuroimage.2010.07.056 (2011).

7       Premoli, I. *et al.* TMS-EEG signatures of GABAergic neurotransmission in the human cortex. *J Neurosci* **34**, 5603-5612, doi:10.1523/jneurosci.5089-13.2014 (2014).

8       Rogasch, N. C. & Fitzgerald, P. B. Assessing cortical network properties using TMS-EEG. *Hum Brain Mapp* **34**, 1652-1669, doi:10.1002/hbm.22016 (2013).

9       Darmani, G. *et al.* Effects of the Selective α5-GABAAR Antagonist S44819 on Excitability in the Human Brain: A TMS-EMG and TMS-EEG Phase I Study. *J Neurosci* **36**, 12312-12320, doi:10.1523/jneurosci.1689-16.2016 (2016).

10      Du, X. *et al.* TMS evoked N100 reflects local GABA and glutamate balance. *Brain Stimul* **11**, 1071-1079, doi:10.1016/j.brs.2018.05.002 (2018).

11      Belardinelli, P. *et al.* TMS-EEG signatures of glutamatergic neurotransmission in human cortex. *Sci Rep* **11**, 8159, doi:10.1038/s41598-021-87533-z (2021).

12      Fecchio, M. *et al.* The spectral features of EEG responses to transcranial magnetic stimulation of the primary motor cortex depend on the amplitude of the motor evoked potentials. *PLoS One* **12**, e0184910 (2017).

13      Tremblay, S. *et al.* Clinical utility and prospective of TMS-EEG. *Clinical neurophysiology : official journal of the International Federation of Clinical Neurophysiology* **130**, 802-844, doi:10.1016/j.clinph.2019.01.001 (2019).

14      Bai, Z., Zhang, J. J. & Fong, K. N. K. Intracortical and intercortical networks in patients after stroke: a concurrent TMS-EEG study. *J Neuroeng Rehabil* **20**, 100 (2023).

15      Rosanova, M. *et al.* Natural frequencies of human corticothalamic circuits. *J Neurosci* **29**, 7679-7685, doi:10.1523/jneurosci.0445-09.2009 (2009).

16      Ferreri, F., Vecchio, F., Ponzo, D., Pasqualetti, P. & Rossini, P. M. Time-varying coupling of EEG oscillations predicts excitability fluctuations in the primary motor cortex as reflected by motor evoked potentials amplitude: an EEG-TMS study. *Hum Brain Mapp* **35**, 1969-1980, doi:10.1002/hbm.22306 (2014).

17      Pellicciari, M. C., Veniero, D. & Miniussi, C. Characterizing the Cortical Oscillatory Response to TMS Pulse. *Front Cell Neurosci* **11**, 38, doi:10.3389/fncel.2017.00038 (2017).

18      Pellicciari, M. C., Veniero, D. & Miniussi, C. Characterizing the Cortical Oscillatory Response to TMS Pulse. *Frontiers in Cellular Neuroscience* **11**, doi:10.3389/fncel.2017.00038 (2017).

19      Dimigen, O. Optimizing the ICA-based removal of ocular EEG artifacts from free viewing experiments. *NeuroImage* **207**, 116117, doi:https://doi.org/10.1016/j.neuroimage.2019.116117 (2020).

20      Farzan, F. *et al.* Characterizing and Modulating Brain Circuitry through Transcranial Magnetic Stimulation Combined with Electroencephalography. *Front Neural Circuits* **10**, 73, doi:10.3389/fncir.2016.00073 (2016).

21      Farrens, J. L., Simmons, A. M., Luck, S. J. & Kappenman, E. S. *Electroencephalogram (EEG) Recording Protocol for Cognitive and Affective Human Neuroscience Research* (Research Square Platform LLC, 2021).

22      Tangwiriyasakul, C. *et al.* Tensor decomposition of TMS-induced EEG oscillations reveals data-driven profiles of antiepileptic drug effects. *Scientific Reports* **9**, 17057, doi:10.1038/s41598-019-53565-9 (2019).

23      Olejarczyk, E. *et al.* Statistical Analysis of Graph-Theoretic Indices to Study EEG-TMS Connectivity in Patients With Depression. *Front Neuroinform* **15**, 651082,

doi:10.3389/fninf.2021.651082 (2021).

24	Farzan, F. *et al.* Reliability of Long-Interval Cortical Inhibition in Healthy Human Subjects: A TMS–EEG Study. *Journal of Neurophysiology* **104**, 1339-1346, doi:10.1152/jn.00279.2010 (2010).

25	Paci, M., Di Cosmo, G., Perrucci, M. G., Ferri, F. & Costantini, M. Cortical silent period reflects individual differences in action stopping performance. *Sci Rep* **11**, 15158, doi:10.1038/s41598-021-94494-w (2021).

26	Cash, R. F. H. *et al.* Characterization of Glutamatergic and GABA<sub>A</sub>-Mediated Neurotransmission in Motor and Dorsolateral Prefrontal Cortex Using Paired-Pulse TMS-EEG. *Neuropsychopharmacology* **42**, 502-511, doi:10.1038/npp.2016.133 (2017).

27	Noda, Y. *et al.* Characterization of the influence of age on GABA(A) and glutamatergic mediated functions in the dorsolateral prefrontal cortex using paired-pulse TMS-EEG. *Aging (Albany NY)* **9**, 556-572, doi:10.18632/aging.101178 (2017).

28	Noda, Y. *et al.* Evaluation of short interval cortical inhibition and intracortical facilitation from the dorsolateral prefrontal cortex in patients with schizophrenia. *Scientific Reports* **7**, 17106, doi:10.1038/s41598-017-17052-3 (2017).

29	Noda, Y. *et al.* A combined TMS-EEG study of short-latency afferent inhibition in the motor and dorsolateral prefrontal cortex. *Journal of Neurophysiology* **116**, 938-948, doi:10.1152/jn.00260.2016 (2016).

30	Zmeykina, E., Mittner, M., Paulus, W. & Turi, Z. Weak rTMS-induced electric fields produce neural entrainment in humans. *Sci Rep* **10**, 11994, doi:10.1038/s41598-020-68687-8 (2020).

31	Ozdemir, R. A. *et al.* Reproducibility of cortical response modulation induced by intermittent and continuous theta-burst stimulation of the human motor cortex. *Brain Stimul* **14**, 949-964, doi:10.1016/j.brs.2021.05.013 (2021).

32	Veniero, D., Ponzo, V. & Koch, G. Paired associative stimulation enforces the communication between interconnected areas. *J Neurosci* **33**, 13773-13783, doi:10.1523/jneurosci.1777-13.2013 (2013).

33	Moffa, A. H. *et al.* Neuromodulatory effects of theta burst stimulation to the prefrontal cortex. *Scientific Data* **9**, 717, doi:10.1038/s41597-022-01820-6 (2022).

34	Moffa, A. H., Nikolin, S., Martin, D., Loo, C. & Boonstra, T. W. Reliability of transcranial magnetic stimulation evoked potentials to detect the effects of theta-burst stimulation of the prefrontal cortex. *Neuroimage: Reports* **2**, 100115, doi:https://doi.org/10.1016/j.ynirp.2022.100115 (2022).

35	Biondi, A. *et al.* Spontaneous and TMS-related EEG changes as new biomarkers to measure anti-epileptic drug effects. *Scientific Reports* **12**, 1919, doi:10.1038/s41598-022-05179-x (2022).

36	Rocchi, L. *et al.* Variability and Predictors of Response to Continuous Theta Burst Stimulation: A TMS-EEG Study. *Frontiers in Neuroscience* **12**, doi:10.3389/fnins.2018.00400 (2018).

37	Delorme, A. & Makeig, S. EEGLAB: an open source toolbox for analysis of single-trial EEG dynamics including independent component analysis. *J Neurosci Methods* **134**, 9-21, doi:10.1016/j.jneumeth.2003.10.009 (2004).